\documentclass[fleqn,usenatbib]{mnras}

\usepackage{newtxtext,newtxmath}
\usepackage[T1]{fontenc}

\DeclareRobustCommand{\VAN}[3]{#2}
\let\VANthebibliography\thebibliography
\def\thebibliography{\DeclareRobustCommand{\VAN}[3]{##3}\VANthebibliography}

\usepackage{graphicx}	% Including figure files
\usepackage{amsmath}	% Advanced maths commands
\usepackage{xfrac}
\usepackage[normalem]{ulem}

\title[Testing the SBF Method on Dwarfs in COSMOS]{Testing the Surface Brightness Fluctuation Method on Dwarf Galaxies in the COSMOS Field}

\author[L. M. Foster et al.]{
Lauren M. Foster,$^{1,2}$
James E. Taylor,$^{1}$
John P. Blakeslee$^{3}$
\\
$^{1}$Department of Physics and Astronomy, University of Waterloo, 200 University Avenue West, Waterloo, Ontario N2L 3G1, Canada\\
$^{2}$Department of Physics and Astronomy, McMaster University, 1280 Main Street West, Hamilton, Ontario L8S 3L8, Canada\\
$^{3}$NSF’s NOIRLab, 950 N. Cherry Avenue, Tucson, Arizona 85719, USA\\
}

\date{Accepted XXX. Received YYY; in original form ZZZ}

\pubyear{2022}

\begin{document}
\label{firstpage}
\pagerange{\pageref{firstpage}--\pageref{lastpage}}
\maketitle

\begin{abstract}
Dwarf galaxies are important tracers of small-scale cosmological structure, yet much of our knowledge about these systems comes from the limited sample of dwarf galaxies within the Local Group. To make a comprehensive inventory of dwarf populations in the local Universe, we require effective methods for deriving distance estimates for large numbers of faint, low surface brightness objects. Here we test the surface brightness fluctuation (SBF) method, traditionally applied to brighter early-type galaxies, on a sample of 20 nearby dwarf galaxies detected in the COSMOS field. These objects are partially resolved in HST ACS images, and have confirmed redshift distances in the range 17--130 Mpc. We discuss the many model choices required in applying the SBF method, and explore how these affect the final distance estimates. Amongst other variations on the method, when applying the SBF method, we alter the standard equation to include a term accounting for the power spectrum of the background, greatly improving our results. For the most robust modelling choices, we find a roughly Gaussian SBF signal that correlates linearly with distance out to distances of 50--100 Mpc, but with only a fraction of the power expected. At larger distances, there is excess power relative to that predicted, probably from undetected point sources. Overall, obtaining accurate SBF distances to faint, irregular galaxies remains challenging, but may yet prove possible with the inclusion of more information about galaxy properties and point source populations, and the use of more advanced techniques. 
\end{abstract}


%%%%%%%%%%%%%%%%%%%%%%%%%%%%%%%%%%%%%%%%%%%%%%%%%%

%%%%%%%%%%%%%%%%% BODY OF PAPER %%%%%%%%%%%%%%%%%%

\section{Introduction}
\label{introduction section}

The Lambda-Cold Dark Matter ($\Lambda$CDM) model underpins modern cosmology, and has proven a great success in predicting the observed abundance and distribution of galaxies on large scales \citep{Mo2010}. On small scales, galaxy formation is strongly modulated by baryonic effects, which clearly suppress the abundance of dwarf galaxies through internal and external feedback processes, although the exact details are unclear and problematic \citep{Bullock2017}. These feedback processes are multiple, and will often depend strongly on local environment, so large statistical samples of dwarfs from a range of different environments are needed to tease them apart.

Unfortunately, a significant part of our observational understanding of dwarf galaxies is limited to the Local Group and its immediate surroundings, where dwarf galaxies can be detected to faint magnitudes and low surface brightness limits, using a variety of techniques \citep{McConnachie2012}. Within the local volume out to 10 Mpc, ongoing efforts are building up extensive samples of satellites \citep[see e.g.][for full lists of references]{Carlsten2022, Nashimoto2022}, but most are in systems similar to the Local Group in mass.
Our census of dwarfs in more and in less massive systems is much less complete, despite the important information these environments provide about the underlying efficiency of galaxy formation \citep[e.g.][Xi \&\  Taylor in prep.]{Sales2013,Grossauer2015,Mueller2020, Garling2021, Roberts2021}. Frustratingly, many of these systems are easily detectable in current wide-area ground based surveys, or will be detected in forthcoming space-based surveys, and their average abundance is also increasingly well constrained \citep{Wang2012,Speller2014,Nierenberg2016, Tanaka2018, Xi2018, Roberts2021, Wang2021, Wu2022}. 
The challenge is to identify exactly which of the huge numbers of objects detected in these surveys are in fact local, without resorting to complete spectroscopic surveys \citep[e.g.][]{SAGA2017}, which are both expensive and challenging, particularly for early-type, low surface brightness dwarfs. This has prompted renewed interest in extending distance estimates based on imaging to fainter populations \citep{Xi2018, Carlsten2019, Polzin_2021, Kim2021, Greco_2021, Carlsten2022}.

%\citet[][X18 hereafter]{Xi2018}\defcitealias{Xi2018}{X18}
\citet{Xi2018}
identified a set of galaxies in the COSMOS field that appear to be partially resolved into point sources and/or surface brightness fluctuations. The objects in the sample are confirmed to be local by their photometric or spectroscopic redshifts. \citet{Polzin_2021} applied the surface brightness fluctuation (SBF -- \citealt{Tonry_1988}) method on the nearest object in the sample (COSMOS ID 549719), and obtained a distance estimate in agreement with the redshift distance. This was an exciting but somewhat surprising result, as the SBF method is normally used on older, brighter elliptical galaxies, whereas they also found the galaxy to have been recently star forming. A natural next step is to test whether this method can produce accurate distance estimates for the larger sample of partially resolved objects compiled by \citet{Xi2018}. 

In this paper, we apply the SBF method to a sample of 20 objects from the \citet{Xi2018} serendipitous catalogue, selected to have spectroscopic redshifts, and redshift-based distances of less than $\sim$130 Mpc. We compare the objects' SBF and proper distances, to explore the limits of the reliability of SBF distance estimates to galaxies of this type at these distances.

The outline of the paper is as follows. In Section \ref{data section}, we describe our sample and the observational data used. In Section \ref{methods section}, we summarize each of the steps in the SBF method. In Section \ref{results section}, we present a fiducial SBF distance estimate for each of the galaxies, following the SBF procedure outlined in Section \ref{methods section}.
In Section \ref{tests section}, we test variations on the fiducial method, and determine their impact on the final results. In Section \ref{discussion section}, we conclude by discussing the limitations of using the SBF method on dwarf galaxies at distances of 20-130 Mpc, and consider how the method might be improved in future work. 

\section{Data}
\label{data section}

\subsection{Galaxy Sample}

The effective range of the SBF technique depends on the depth of the imaging used, but for bright galaxies imaged with HST, reaches 100-130\,Mpc \citep{Blakeslee2021, Moresco2022}. For faint, low surface brightness galaxies, the range is unclear, but
\citet{Xi2018} found a reasonable correspondence between general optical morphology (i.e.~visual indications that systems were partially resolved) 
and distance for dwarf galaxies with $i^+ < 20$ mag, out to distances of $\sim$ 200\,Mpc. To test the effective range of the technique for fainter, less regular galaxies, we considered the 34 objects in their serendipitous catalogue with spectroscopic redshifts that put them within $\sim$130\,Mpc.

Of these, 14 galaxies were excluded from our final sample;
5 because they were clearly intrinsically bright and/or star-forming galaxies,
2 because they had an extremely irregular morphology that would make model fitting difficult (see \ref{imfit model section}), and 7 because they were too small, faint, or low surface brightness. The COSMOS2015 IDs, heliocentric redshift distances (d = $cz/H_0$, assuming $H_0 = 70 {\rm\, km\, s^{-1}Mpc^{-1}}$), and coordinates of the remaining 20 objects are listed in Table \ref{data table}, along with the cutout image sizes used in our subsequent analysis.

\subsection{Imaging and Spectroscopy}

Each of the galaxies in our final sample has 
multi-band photometry from the COSMOS 2015 catalogue \citep{Laigle_2016}, as well as a spectroscopic redshift given in \citet{Xi2018}. We used these catalogue redshifts in all cases except for 549719, where \cite{Polzin_2021} measured a redshift of 1222 $\pm 64$ km\,s$^{-1}$, and 213165, whose catalogue redshift places the galaxy too close. For this galaxy, a corrected redshift was obtained from the NASA Extragalactic Database (NED)\footnote{https://ned.ipac.caltech.edu}. 

The HST images used in the SBF analysis were HST/ACS F814W-band mosaics \citep{Koekemoer_2007, Massey_2010} obtained from the NASA/IPAC IRSA COSMOS Cutouts Service, 
while Subaru $i^+$, IA464, and IA484-band images \citep{Taniguchi_2015} from the same source were used to estimate the $g-i$ colour of the galaxies.

\begin{table}
    \centering
    \caption{COSMOS galaxy ID, redshift distance, coordinates, and image size}
    \begin{tabular}{c | c | c | c | c}
    
    COSMOS2015 & Redshift & Right & & Image \\
    ID & Distance & Ascension & Declination & Size  \\
     & (Mpc) & (deg) & (deg) &  ($^{\prime\prime}$) \\\hline
    
    549719 & 17.5  & 150.12526 & 2.14985 & 18 \\ % DG1
    424575 & 21.5  & 149.51290 & 1.95290 & 25 \\ % NDGX1
    677414 & 24.8  & 149.69517 & 2.34761 & 40 \\ % DG2
    260583 & 26.0  & 149.62025 & 1.69362 & 80 \\ % DG3
    561851 & 26.6  & 150.61308 & 2.16682 & 15 \\ % DG7
    401988 & 27.0  & 150.02440 & 1.91103 & 45 \\ % DG18
    733922 & 29.0  & 150.47407 & 2.41385 & 30 \\ % DG21
    686606 & 31.0  & 150.36668 & 2.34040 & 20 \\ % DG22
    259971 & 43.0  & 149.46120 & 1.67490 & 15 \\ % DG16
    213165 & 43.0  & 150.69502 & 1.61378 & 25 \\ % DG12
    709026 & 50.6  & 150.02833 & 2.37924 & 15 \\ % DG6
    918161 & 51.6  & 150.39213 & 2.69175 & 15 \\ % DMAR9
    458976 & 57.6  & 149.86624 & 2.00702 & 20 \\ % DG23
    331749 & 81.7  & 150.34559 & 1.79364 & 20 \\ % DG19
    880547 & 103.2 & 150.00229 & 2.63321 & 15 \\ % DG5
    380820 & 103.2 & 150.06000 & 1.86660 & 15 \\ % DMAR4
    589205 & 107.5 & 149.81181 & 2.19223 & 15 \\ % DG24
    279307 & 109.2 & 149.96432 & 1.70670 & 20 \\ % DG11
    377112 & 116.1 & 150.19167 & 1.86347 & 20 \\ % DST6
    824852 & 125.1 & 149.75697 & 2.54992 & 15 \\ % NDST2
    \end{tabular}
    \label{data table}
\end{table}

\section{Method}
\label{methods section}

If the light emitted by a galaxy were produced entirely by stellar point sources of fixed luminosity, we would expect the number $N$ of these sources falling within a single pixel of the image to scale with the physical area subtended by the pixel, and thus to vary as the square of the geometric distance to the galaxy. Poisson fluctuations in this number from pixel to pixel, across a region of uniform brightness, should vary as $N^{1/2}$, that is as $D$. For more realistic stellar populations with a range of intrinsic luminosity, the amplitude of fluctuations should depend on a weighted moment of the luminosity function, and thus on the age and metallicity of the population, and the resulting distribution would no longer be purely Poissonian \citep{Cervino2006, Cervino2008}. Given an absolute calibration for a similar stellar population, however, one should still be able to infer the distance from the amplitude of pixel-to-pixel variations, relative to the mean surface brightness. Based on this idea, \citet{Tonry_1988} first proposed the Surface Brightness Fluctuation (SBF) method for determining absolute distances to nearby galaxies. 

In practice, the SBF method requires a number of distinct steps, including masking individual point sources and foreground/background objects, modelling and subtracting and/or dividing by the average light distribution across the galaxy, measuring the spatial power spectrum of the residual map, the background sky around the object, and the point spread function (PSF) of the image, fitting the overall power spectrum to these components, and adjusting the final result, cast as an `SBF magnitude', for an age and metallicity-dependent zero-point. We discuss these steps, and the possible choices to make at each, below.

\subsection{Masking and Modelling}
\label{masking and modelling section}

An important first step in the SBF method is to mask any potential contaminants in the image that could give rise to extra fluctuations unrelated to discreteness noise in the underlying stellar population. Examples include foreground stars, background galaxies, globular clusters, and cosmetic artifacts in the image. Once the image is masked, we can create a smooth model of the galaxy's light distribution, to be used in subsequent steps. Masking and modelling are interconnected and happen simultaneously, so we will discuss them both in this section.

We use three methods for masking the image to identify bad regions and exclude them from subsequent analysis: manual masking, masking with automatic point source detection, and masking using the galaxy image pixel histogram. Each of the three methods builds on and incorporates the previous one(s). 

\subsubsection{Initial Manual Masking}
\label{manual mask section}
We first masked the images manually, by inspecting them visually, noting which areas of the image contained bad pixels or contaminating objects, and masking them as simple circular or rectangular regions. The goal of this process was to remove the most obvious contaminating sources from each image, so that the subsequent steps were more efficient.

\subsubsection{Sérsic Model}
\label{imfit model section}
Given our initial masking, the next step was to model the smooth light distribution of the galaxy. To do this, we initially created elliptical Sérsic models of each galaxy, using \texttt{imfit} \citep{Erwin_2015}, with our manual mask indicating regions to ignore, and the absolute value of the most negative pixel in the image taken to be the value of the previously subtracted constant sky background. We used reasonable guesses for the initial parameters of the fit in each case, including $n=0.5$, $I(r_e)=0.01$ counts/s, and $r_e=100$ pixels, but found these had little impact on the final fitted parameters (see table \ref{property table}), as did the constraints we put on parameters. To determine errors on the fit, we used \texttt{imfit}'s bootstrap resampling command, with 200 iterations. Given the Sérsic Model fit, we restricted our subsequent SBF analysis to an elliptical region centred on the galaxy with a semi-major axis of $2r_e$, and the same axis ratio as the Sérsic Model.

\subsubsection{Bicubic Spline Model}
\label{bicubic spline model section}
We also created a second, non-parametric model for the smooth light distribution of the galaxy using bicubic spline interpolation \citep{Jordan2004}. First, we made a lower resolution version of the masked image, with each 40x40 pixel area on the original image corresponding to a single pixel on the lower resolution image. Then, we fit a bicubic spline function to the logarithm of the lower resolution version of the image using \texttt{scipy.interpolate.griddata}. Finally, we created the model by calling the bicubic spline function at the original resolution grid spacing, and taking the reverse logarithm of the resulting image.

To account for the mask affecting adjacent pixels during the process of lowering the resolution and interpolating between pixels, we also applied the spline fitting process to the manual mask. We masked all pixels in the resulting `mask model' that had values less than 0.9, in effect expanding the edges of the manual mask.

\subsubsection{Automatic Point Source Mask}
\label{auto mask section}
Given our manual mask and smooth model for the light distribution in each galaxy, we then used the \texttt{photutils.detection.IRAFStarFinder} method from the \texttt{photutils} package \citep{larry_bradley_2022_6825092} to automatically detect any point sources in the images that had been missed when masking manually,
choosing a threshold of 0.018 counts/s, a background of 0, and a PSF FWHM of 3 pixels.
For each galaxy, we masked each detected source using a circular area with a radius of 6 pixels. For efficiency, we only searched for point sources within the $2r_e$ region of interest defined above. Point source properties are described further in Appendix \ref{luminosity function appendix}.

\subsubsection{Smoothed Image Model}
\label{smoothed image model section}
We then created a third, non-parametric model for the smooth light distribution in each galaxy, by simply convolving the masked image with a Gaussian kernel with a 5 pixel radius.
This empirical model has the benefit of more accurately capturing any irregular structure in the galaxies, and being more customized to each individual galaxy. 

\subsubsection{Histogram Mask}
\label{histogram mask section}
\begin{figure*}
    \centering
    \includegraphics[width=\linewidth]{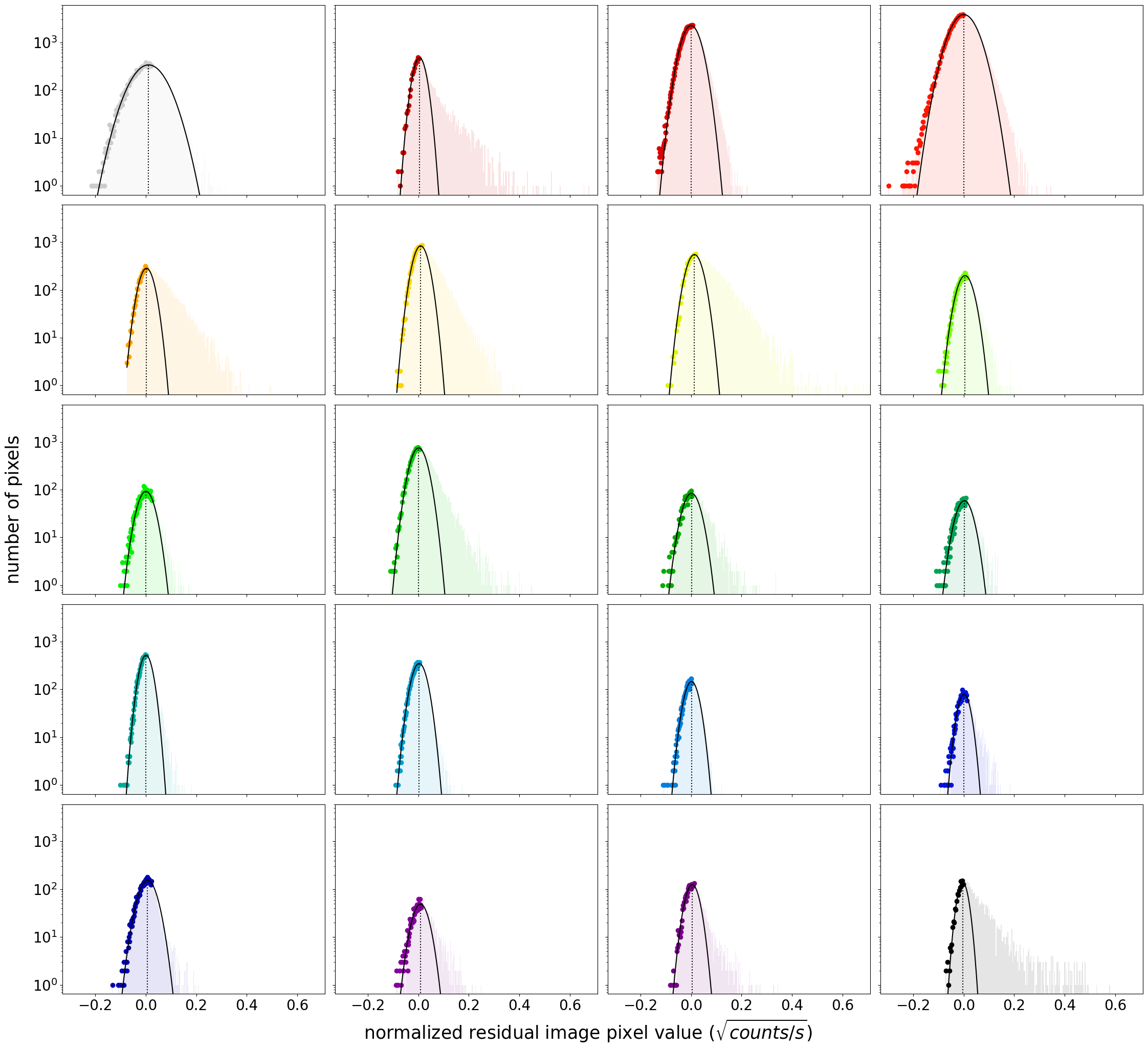}
    \caption{Pixel value distributions in the Normalized Residual Images (NRIs -- coloured histograms) created using the smoothed image model. Note that in this and most subsequent plots, the sample is ordered by distance from top left to bottom right. The thick points show the bins used to fit the {\color{black}Gaussian} component, while the black curves and vertical lines show the final fit and fitted peak value respectively. The histogram mask excludes pixels with values in excess of the {\color{black}Gaussian} expectation (i.e.~the tail to the right of the Gaussian component). \textcolor{black}{Galaxies are ordered from left to right as follows: 424575, 549719, 677414, 260583, 561851, 401988, 733922, 686606, 213165, 259971, 709026, 918161, 458976, 331749, 880547, 380820, 589205, 279307, 377112, 824852.}}
    \label{NRI histogram figure}
\end{figure*}

Finally, we constructed a third, more sophisticated mask based on a pixel histogram of the galaxy's normalized residual image (NRI; see Section \ref{power spectrum section} for details on the construction of this image), masked with the automatic detection mask. The pixel histogram of the residual image (Fig.~\ref{NRI histogram figure}) includes a Gaussian component at low counts/s, but in some cases also an excess tail at high counts/s. {\color{black} We expect intrinsic surface brightness fluctuations to be Gaussian in the limit where large numbers of stars contribute significantly to the light in each pixel. Since a Gaussian component appears in the NRI distributions for all the objects in the sample, whereas only a few show a significant high-residual tail, for this masking method we assume the Gaussian component corresponds to intrinsic surface brightness fluctuations, whereas the rest of the distribution is produced by other effects. We mask any pixels with residuals in the range where the NRI histogram greatly exceeds the Gaussian component, as follows.} 

We create the histogram mask by estimating the peak of the distribution to be the bin with the most pixels, $x_{max}$, and then fitting a normal distribution to a region around and below the peak, below a threshold value:
\begin{equation}
    x_{\rm threshold} = x_{\rm max} + |0.2x_{\rm max}|\,.
    \label{histogram threshold equation}
\end{equation}
This ensures that the fit is not affected by the non-{\color{black}Gaussian} tail. If, due to noise, the fitted value of the peak is above $x_{\rm threshold}$, we iterate until it drops below this value.

Given our fit to the {\color{black}Gaussian} component of the pixel distribution, we then mask any pixels beyond the point where the fit intercepts the line $y=1$, i.e., where we would only expect one pixel in the bin with this pixel value. This produces the most aggressive of our three masks.

\textcolor{black}{We note that the galaxies in our sample have on the order of $10^3M_\odot$ per pixel. From \citep{Tonry_1988}, and the results of \citep{Cervino2006, Cervino2008}, this seems too low to expect fully Gaussian fluctuations, so it may be that the high-residual tails seen in the NRI for 6--7 of our galaxies are genuinely due to stellar discreteness noise, and the corresponding pixels should be included in the SBF calculation. On the other hand, the fact they generally appear in objects with {\it larger} stellar surface mass density makes this seem less likely. As discussed in appendix \ref{luminosity function appendix}
, these galaxies also have the largest number of bright point sources, so residuals in the point source subtraction may explain the presence of the tails.}

%Figure 1

\subsection{Computing the Power Spectrum}
\label{power spectrum section}

The next step in the SBF method is to calculate the normalized residual image, and determine what fraction of the pixel-to-pixel variation in this image is due to SBFs. In practice, this is usually done in Fourier space, since for a raw CCD image, the power spectrum of the image should include an `on-sky' component that is convolved with the PSF, and a white noise component from the readout electronics \citep{Blakeslee1999}. 

After masking and modelling the galaxy, we create a normalized residual image (NRI). The NRI is calculated as:
\begin{equation}
    NRI = \frac{image-model}{\sqrt{model}}\,,
    \label{NRI equation}
\end{equation}
where we first subtract the model from the image, so that only the fluctuations remain, and then divide by the square root of the model, to normalize the scale of the fluctuations to the expected value. 

To compute the power spectrum of the NRI, we create an elliptical aperture around the galaxy and combine this with the mask, to ensure that any measured SBF variance is coming from the galaxy rather than from other sources. 

Since SBFs occur in the real image of the galaxy on the sky, they will be convolved with the telescope's PSF. Taking the Fourier transform of the image converts this convolution to a multiplication, so we can simply divide the power spectrum of the NRI, minus any white noise component due to readout noise, by the power spectrum of the PSF, to estimate the amplitude of the SBFs. 

Two complications arise. First, COSMOS mosaic images have been combined from multiple raw exposures, using drizzling with an interpolation kernel for geometric corrections, so the white noise component is not actually `white', but has a slight dependence on wavenumber (as discussed further in section \ref{background tests section}; see also \citealt{Mei2005}). To correct for this, we estimated the power spectrum of the background from blank regions in the image around the galaxy, and subtracted this empirical form when fitting for the SBFs. 

A second complication is masking. This has a multiplicative effect on the image in real space, and thus corresponds to a convolution in frequency space. To account for this, we multiply the SBF component of the power spectrum by a Fourier transform of the PSF, convolved with a Fourier transform of the full mask (taken to be the product of the manual mask, the automatic point source mask, the histogram-based mask, and the elliptical aperture described above). All Fourier transforms were computed in 2D using the Python function \texttt{numpy.fft.fft2}; azimuthal averages were then taken for each bin in wavenumber.

\begin{figure}
    \centering
    \includegraphics[width=\linewidth]{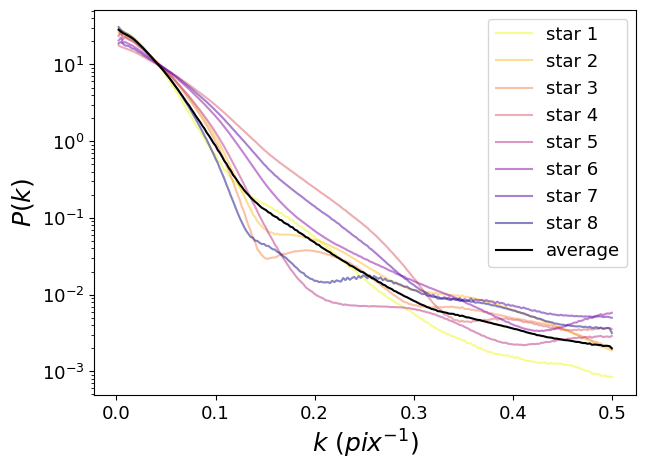}
    \caption{Power spectra of the eight non-saturated stars used to estimate the PSF (thin coloured curves), as well as the power spectrum of the mean PSF (thick black curve).}
    \label{average PSF figure}
\end{figure}

To compute the PSF, we used eight non-saturated stars from across the COSMOS field. We took circular areas of radius 25 pixels around each star and masked the rest of the image. We then computed the 2D power spectrum of each star and took the average of all eight power spectra as the mean power spectrum of the PSF. Fig.~\ref{average PSF figure} shows the resulting 1D power spectra of the eight individual stars, as well as the mean PSF.

%Figure 2

\begin{figure}
    \centering
    \includegraphics[width=\linewidth]{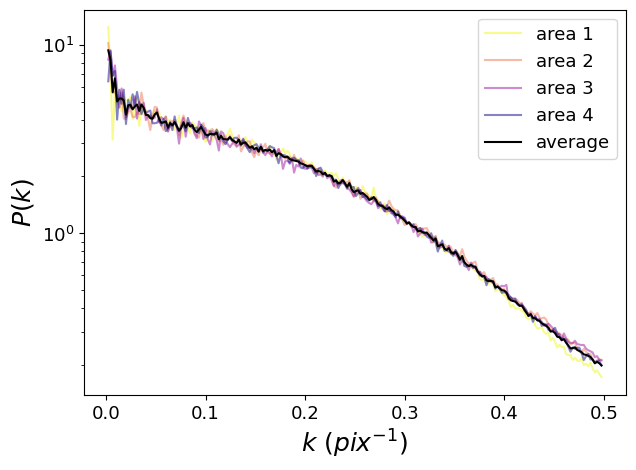}
    \caption{Power spectra of the four empty regions used to estimate the power spectrum of the background (thin coloured curves), as well as the mean power spectrum of the background (thick black curve).}
    \label{average background figure}
\end{figure}

To compute the power spectrum of the background, we used four nearly empty  areas (i.e.~with only minimal masking required) in the COSMOS field, ranging in size from 15--20$\arcsec$ \textcolor{black}{(500-667 pixels)} on a side. We modelled each area using \texttt{imfit}'s FlatSky function, which only has one parameter, the surface brightness of the sky, to calculate a normalized residual image for each. Calculating the power spectrum of each and normalizing it to have an integral of 1, we then interpolated the resulting background spectrum to have the same sampling in $k$ as each individual galaxy image. Finally, we took the mean of the four normalized and interpolated spectra to get the power spectrum of the background for each galaxy. Fig.~\ref{average background figure} shows the resulting normalized 1D power spectrum of the background.

%Figure 3

\subsubsection{Noise Estimate}

We have found experimentally that the shape and amplitude of the NRI power spectrum change slightly with the choice of aperture size and position, so to avoid creating any bias from a particular choice of these parameters, we created 50 different power spectra for each galaxy, randomly selecting the size of the semi-major axis $a$ and the position of the centre $(x_0,y_0)$ from uniform distributions: $r_e\leq a<2r_e$, and $x_0-5\leq x<x_0+5$, $y_0-5\leq y<y_0+5$, where $r_e$ is the effective radius from \texttt{imfit} and $x_0$ and $y_0$ are the coordinates of the aperture centre in pixels. We held the ellipticity and angle constant, using the best-fit \texttt{imfit} values for these parameters.

\subsection{Estimating the SBF Variance}
\label{SBF variance section}

The SBF variance is usually calculated by fitting the azimuthally-averaged power spectrum of the masked NRI, $P(k)$, to the equation \textcolor{black}{\citep{Tonry1990}}:
\begin{equation} 
    P(k)=P_0 \times E(k) + P_1
    \label{SBF original equation}
\end{equation}
where $P_0$, the quantity of interest, is the SBF variance, $P_1$ is the white noise variance, and $k$ is the spatial frequency in units of inverse pixels \citep[e.g.][]{Greco_2021}. $E(k)$ is known as the (azimuthally-averaged) expectation power spectrum, and is a 1D average of the two-dimensional convolution of the power spectra of the PSF and the mask \citep{Greco_2021}:
\begin{equation}
    E(k_x,k_y)=|\text{PSF}(k_x,k_y)|^2*|M(k_x,k_y)|^2\,.
    \label{expectation power spectrum equation}
\end{equation}
 We used \texttt{astropy.convolution.convolve\textunderscore fft} to compute the expectation power spectrum, given the power spectra of the PSF and the mask, and normalized it to have an integral of 1 before fitting.

As mentioned above, however, we found that the white-noise component of the spectrum actually has a dependence on $k$ (see Section \ref{background tests section}). To account for this, we modified equation (\ref{SBF original equation}) to include the power spectrum of the background variance:
\begin{equation} 
    P(k)=P_0 \times E(k) + P_1 \times B(k)
    \label{SBF background equation}
\end{equation}
where $E(k)$ and $B(k)$ are the expectation power spectrum and the power spectrum of the background respectively, computed as described above.

We fit to equation (\ref{SBF background equation}) using \texttt{scipy.optimize.curve\textunderscore fit}, which uses a non-linear least squares algorithm. Given the model from equation (\ref{SBF background equation}), it returns the best fitting $P_0$ and $P_1$ values and the covariance for these. \textcolor{black}{Note that when fitting, we use units of inverse pixels for $k$ for convenience, i.e.~we divide the standard wavenumber $k$ by the number of pixels on one side of the image.}

In fitting the expectation power spectrum, we limit the range of wavenumbers considered. Very low wavenumbers (large spatial scales) are affected by the smoothing used to make the empirical model of the smooth light distribution, while high wavenumbers are compromised by correlated noise between pixels \citep{Greco_2021} and/or errors in our background variance estimation. 
As discussed in Section \ref{wavenumber tests section}, we tested many wavenumber ranges for the entire sample, finding the range $0.1$--$0.3\,\text{pix}^{-1}$ to be optimal.

\subsection{Computing the Fluctuation Magnitude and the SBF Distance}
\label{SBF magnitude and distance section}

Each of the 50 SBF runs for a single galaxy returns a value of $P_0$, which needs to be divided by the number of unmasked pixels, $N$, as $P_0$ is the sum of pixel variances \citep{Tonry_1988, Greco_2021}. We computed the mean and standard deviation of the 50 $P_0/N$ values rather than the mean of the SBF distances to estimate the uncertainty because some of the mean $P_0/N$ values are negative, mapping to an infinite distance. \textcolor{black}{This can happen when the data is noisy and the background term dominates equation (\ref{SBF background equation}).} We calculated the apparent fluctuation magnitudes from: 
\begin{equation}
    m_{\rm SBF}=m_{\rm zpt}-2.5\log_{10}{\frac{P_0}{N}}\,,
    \label{SBF apparent magnitude equation}
\end{equation}
where $m_{\rm zpt}$ is the zero point magnitude of the original image.

To convert the apparent fluctuation magnitude to a distance, we need an independent estimate of the calibration magnitude $M_{\rm SBF}$. We used the calibration developed by \cite{Carlsten2019}, that was calculated using tip of the red-giant branch distances for low surface-brightness dwarf galaxies. $M_{\rm SBF}$ is calculated using the colour of the galaxy, because the amplitude of SBF is dependant on the stellar luminosity function of the galaxy, and thus on its stellar population. The age and metallicity of the population can be accounted for using an optical colour, thus the net dependence of $M_{\rm SBF}$ on colour \citep{Worthey1994, Carlsten2019, Polzin_2021}. 

We tested two separate methods of computing the $g-i$ colour needed for the calibration, calculating this either from the COSMOS photometry, or from the images directly. See Section \ref{colour tests section} for a comparison of the two methods. For the first method, we estimated the $g-i$ value using the COSMOS 2015 catalogue's 3$\arcsec$ \textcolor{black}{(corresponding to 100 pixels in the F814W imaging)} aperture magnitudes in the Subaru bands $i^+$, IA464, and IA484. This aperture was selected, as the automatic aperture magnitudes give noisier colours \citep{Laigle_2016}. The g-band magnitude is not available, so we estimated it as:
\begin{equation}
    m_g \approx \frac{m_{IA484}+m_{IA464}}{2}\,.
    \label{g band equation}
\end{equation}
\begin{equation}
    g-i \approx m_g(3\arcsec) - m_i(3\arcsec)
    \label{colour equation}
\end{equation}

The second method involves the same $g-i$ estimation but using the Subaru $i^+$, IA464, and IA484 images directly. For each galaxy, we created an approximate g-band image using equation (\ref{g band equation}), and used this to make  $g-i$ image. We then applied to this colour image an elliptical aperture identical to that used on the galaxy. We took the average of all the values within the aperture as the estimated $g-i$ colour, and the standard deviation as its error.

To use the \cite{Carlsten2019} calibration, we converted the galaxy's $g-i$ colour to a F475W-F814W colour by solving \textcolor{black}{equation (3) in \cite{Carlsten2019}:}
\begin{equation}
    g-i=x-0.061x^2-0.040x-0.003
    \label{convert colour}
\end{equation}
where $x$ is the F475W-F814W colour. Next, we used \textcolor{black}{equation (4) in \cite{Carlsten2019}:}
\begin{equation}
    M_{SBF,i}=(-3.17\pm0.19)+(2.15\pm0.35)\times(g-i)
    \label{Msbf i}   
\end{equation}
to calculate the SBF absolute magnitude in the $i$-band. Since our imaging is in the F814W band, however, this was converted to this band using \textcolor{black}{equation (2) in \cite{Carlsten2019}:}
\begin{equation}
    M_{SBF,814}=M_{SBF,i}-0.540x^2+0.606x-0.253\,.
\end{equation}

Finally, the SBF distance was calculated using:
\begin{equation}
    m_{\rm SBF}-M_{\rm SBF}=5\log_{10}{\frac{d}{10\text{pc}}}
    \label{distance modulus}
\end{equation}

\section{Results}
\label{results section}

The derived SBF distances were compared to distances based on the redshifts described in Section \ref{data section}. To correct the latter for the effects of nearby structure, including the Virgo Cluster, the Great Attractor, and the Shapley Supercluster, we also derived flow-corrected proper distances based on the model of \citep{cosmicflows}, by looking up the CMB-frame velocity in NED and then using the on-line tool at http://cosmicflows.iap.fr, with the cosmological parameters $(\Omega_{\rm m}, H_0) = (0.315, 67.5 {\rm\, km\, s^{-1}Mpc^{-1}})$. (Note CMB-frame velocities are typically $\sim300$ km\, s$^{-1}$ larger than heliocentric velocities in the COSMOS field.)
We assigned all flow-corrected distances an error of 4.3 Mpc, corresponding to $300$ km\, s$^{-1}$ for an $H_0$ value of $70 {\rm\, km\, s^{-1}Mpc^{-1}}$, to account for peculiar velocities. We use the proper distances given in Table \ref{redshift table}, for all subsequent comparisons.

\begin{table}
    \centering
    \caption{COSMOS galaxy ID, redshift distance, and estimated proper distance}
    \begin{tabular}{c | c | c}
    
    COSMOS & Redshift & Proper \\
    ID & Distance (Mpc) & Distance (Mpc) \\\hline
    
    549719 & 17.5  & 21.3  \\ % DG1
    424575 & 21.5  & 27.5  \\ % NDGX1
    677414 & 24.8  & 31.8  \\ % DG2
    260583 & 26.0  & 33.3  \\ % DG3
    561851 & 26.6  & 33.4  \\ % DG7
    401988 & 27.0  & 34.2  \\ % DG18
    733922 & 29.0  & 37.2  \\ % DG21
    686606 & 31.0  & 37.6  \\ % DG22
    259971 & 43.0  & 50.8  \\ % DG16
    213165 & 43.0  & 50.8  \\ % DG12
    709026 & 50.6  & 58.2  \\ % DG6
    918161 & 51.6  & 59.1  \\ % DMAR9
    458976 & 57.6  & 64.0  \\ % DG23
    331749 & 81.7  & 88.5  \\ % DG19
    880547 & 103.2 & 112.9 \\ % DG5
    380820 & 103.2 & 115.0 \\ % DMAR4
    589205 & 107.5 & 115.8 \\ % DG24
    279307 & 109.2 & 116.7 \\ % DG11
    377112 & 116.1 & 126.0 \\ % DST6
    824852 & 125.1 & 133.6 \\ % NDST2
    \end{tabular}
    \label{redshift table}
\end{table}

Our fiducial SBF distances assume the following choices or parameters:
\begin{itemize}
    \item the smoothed image model to compute the NRI
    \item the histogram mask (incorporating the preceding manual and automatic masks) to remove contaminating sources and bad regions
    \item the realistic background SBF equation (equation (\ref{SBF background equation})) for background correction
    \item a $g-i$ colour derived directly from the image
    \item a wavenumber range of $0.1$ to $0.3\,\text{pix}^{-1}$ when fitting the power spectrum\,.
\end{itemize}
Also, we run the SBF method 50 times on each object, selecting a random aperture each time (see Section \ref{power spectrum section}), to estimate the uncertainties in $P_0$ and $P_1$, and propagate the former through the distance calculations. 

\begin{figure}
    \centering
    \includegraphics[width=\linewidth]{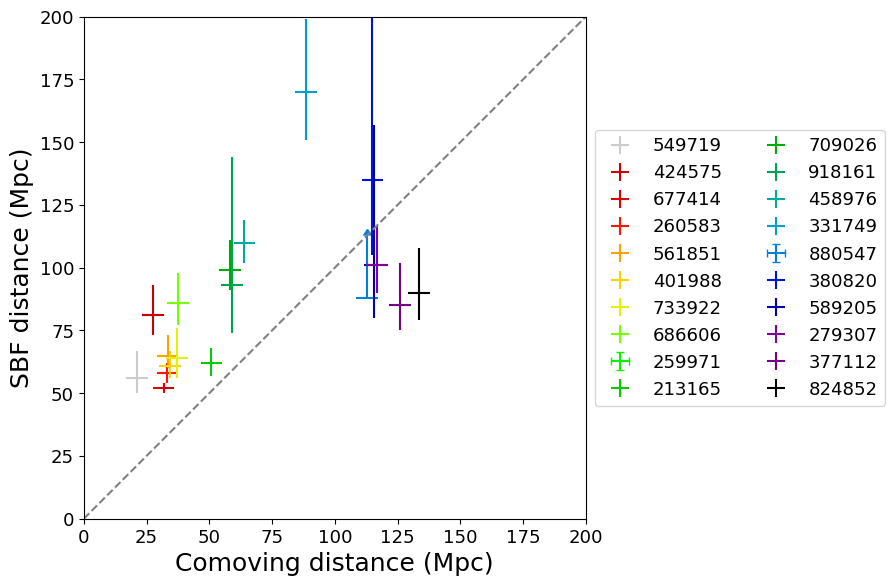}
    \caption{SBF distance $d_{\rm SBF}$ versus proper distance $d_{c}$. Objects with upper and lower limits are plotted with vertical error bars corresponding to $\pm1\sigma$ in $P_0/N$. Objects with a finite lower limit but an infinite upper limit are plotted as upward-pointing arrows. Note the labels in the legend are sorted by increasing distance, as in Fig.~\ref{NRI histogram figure}.} 
    \label{dsbf vs dz figure}
\end{figure}

The results are shown in Fig.~\ref{dsbf vs dz figure}, where we plot SBF distances versus proper distances, with the dashed grey line representing a 1:1 correspondence. Error bars are derived from the set of 50 variants on the fiducial aperture described above \textcolor{black}{($\pm1\sigma$ in $P_0/N$)}, and probably underestimate the true errors; these are hard to define precisely, as they depend on the range of choices we are willing to make in the method.

%Figure 4

\begin{figure}
    \centering
    \includegraphics[width=\linewidth]{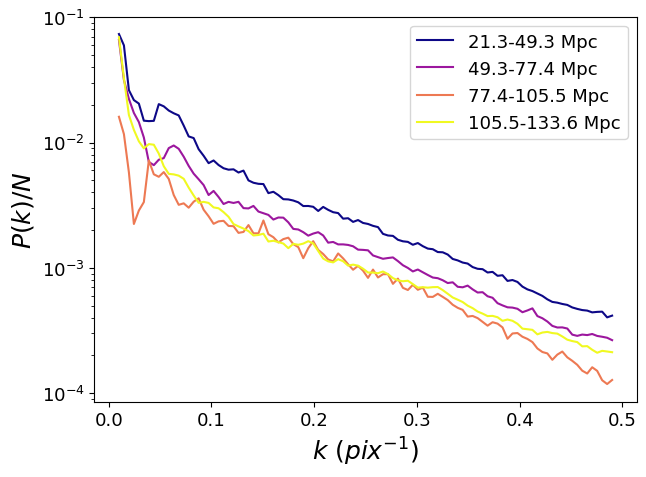}
    \caption{Normalized NRI power spectra $P(k)/N$, grouped by proper distance and averaged.}
    \label{grouped power spectra figure}
\end{figure}

\begin{figure}
    \centering
    \includegraphics[width=\linewidth]{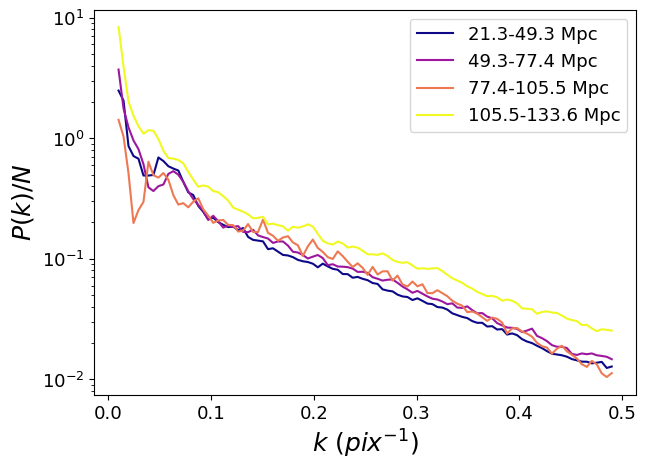}
    \caption{Average NRI power spectra for each of the four bins in proper distance, with the individual power spectra multiplied by their distance before averaging.}
    \label{grouped power spectra times distance figure}
\end{figure}

For galaxies at flow-corrected distances of less than 100 Mpc, there is a clear correspondence between the SBF fluctuation amplitude and actual distance. Fig.~\ref{grouped power spectra figure} shows the mean power spectra for galaxies grouped into four bins by proper distances: 21.3--49.3 Mpc (8 objects), 49.3--77.4 Mpc (5 objects), 77.4--105.5 Mpc (1 object), and 105.5--133.6 Mpc (6 objects). Individual power spectra (within $1r_e$ aperture) were normalized by the number of unmasked pixels and averaged in each bin. We see a clear trend of decreasing power with distance, except in the fourth bin, which has a mean power spectrum similar to the third. Since we expect $P_0/N$ to scale linearly with distance, we can normalize the results by distance by multiplying the power spectrum for each object by its distance before averaging; once again, Fig.~\ref{grouped power spectra times distance figure} shows that the first three bins have comparable intrinsic power, while the last bin does not.

%Figure 5

%Figure 6

\begin{figure}
    \centering
    \includegraphics[width=\linewidth]{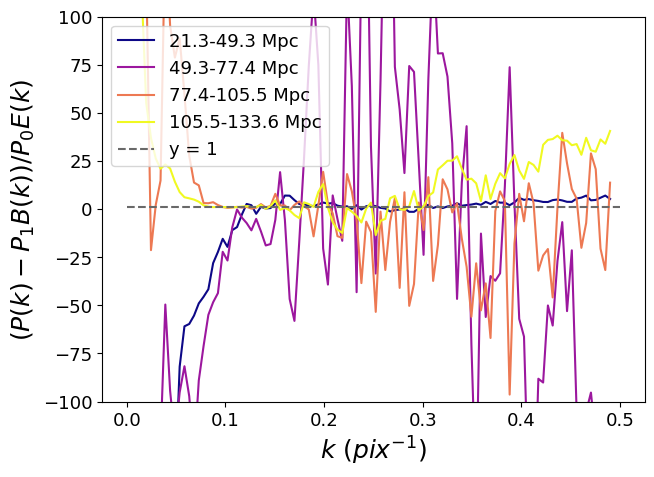}
    \caption{Residuals of the average NRI power spectra for each of the four bins in proper distance, after subtracting the fitted background component and dividing by the fitted SBF component. The dashed line indicates the theoretical expectation, $y=1$.}
    \label{grouped power spectra residual divided by expectation figure}
\end{figure}

Finally, Fig.~\ref{grouped power spectra residual divided by expectation figure} shows residual NRI power spectra, after subtracting the fitted background component and dividing by the fitted SBF/Mask power spectrum:
\begin{equation}
    \frac{P(k)-P_1B(k)}{P_0E(k)}\,.
    \label{residual divided equation}
\end{equation}
As before, the results for each galaxy have been averaged in four distance bins. If our modelling and fit are correct, the residuals should be constant with $y=1$ (black dashed line). This is appears to be somewhat true in the range $k=0.13$ -- $0.33$, but outside this range we see significant positive and negative variations. (We also note that the third bin is very noisy because it contains only one object.)

%Figure 7

From all this, we conclude that SBF power in our sample scales as expected out to $\sim$60--100 Mpc, but that beyond that range, some other component that is independent of distance dominates the power spectrum. At smaller distances, there is an additional problem with the zero-point of the relation. All of our sample points lie above the 1:1 line in Fig.~\ref{dsbf vs dz figure}, indicating that we are systematically underestimating the SBF power in all of them. 

\begin{figure}
    \centering
    \includegraphics[width=\linewidth]{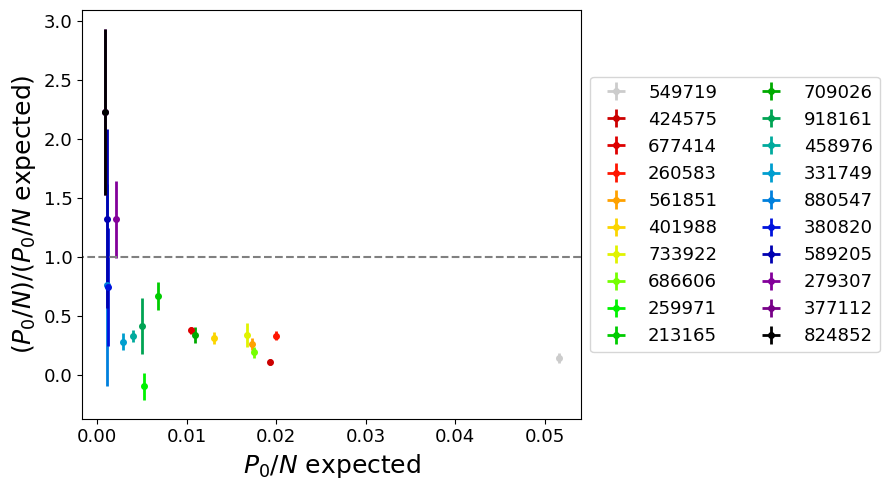}
    \caption{Measured SBF power $P_0/N$ versus expected power, as a function of expected power.}
    \label{Power_vs_expected}
\end{figure}

Fig.~\ref{Power_vs_expected} illustrates these two problems for individual galaxies; distant galaxies generally have more power than expected, while nearby galaxies have only a third the power expected. We will explore these problems further in Section \ref{tests section}.

%Figure 8

\section{Testing Variations on the SBF Method}
\label{tests section}

There are a number of choices to make when proceeding through the steps in the SBF method. They include the type of smooth model used, the type of mask used, how the background contribution is subtracted, the size of the aperture applied, the range of wavenumber values to fit, and the optical colour to use in the calibration. In this section, we discuss how we made each of these choices, and show how variations on our fiducial choices affect the results.

\subsection{Modelling the Smooth Light Distribution}
\label{model tests section}
In Section \ref{masking and modelling section}, we described three different models of the smooth light distribution in the galaxy: a Sérsic profile created using \texttt{imfit}, a bicubic spline model, and a smoothed-image model created by convolving the image with a Gaussian kernel. The model of the light distribution is used when creating the NRI, using equation (\ref{NRI equation}). 

\begin{figure}
    \centering
    \includegraphics[width=\linewidth]{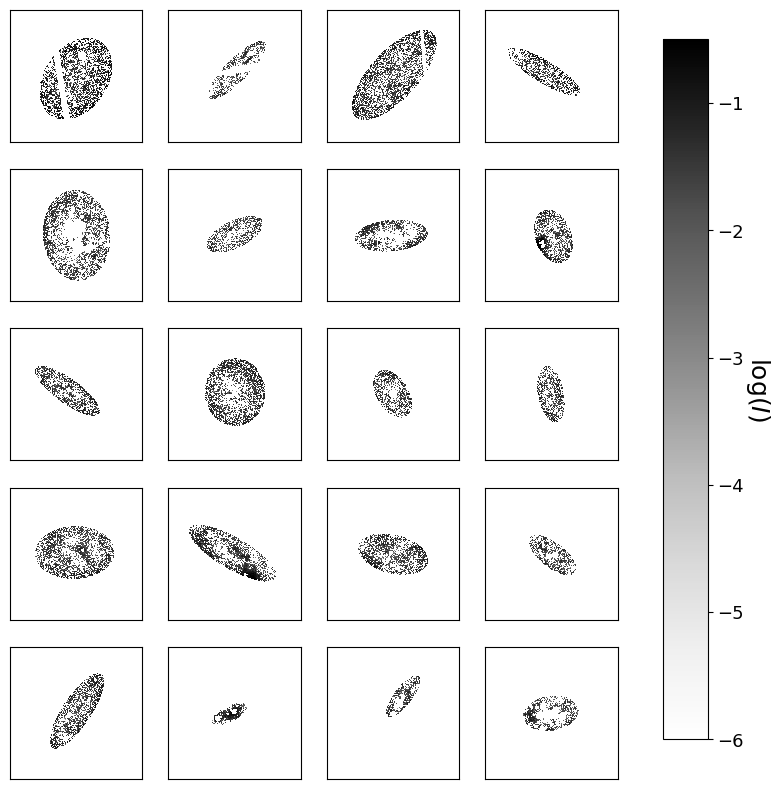}
    \caption{Normalized residual images created using the \texttt{imfit} models. Each image has a $2r_e$ aperture applied around it, to focus on the residuals left behind after fitting. \textcolor{black}{Galaxies are ordered as in Fig.~\ref{NRI histogram figure}.}}
    \label{imfit NRIs figure}
\end{figure}

Using the parametric \texttt{imfit} model, we found that the NRI contains a lot of large-scale residual structure (Fig.~\ref{imfit NRIs figure}). This is because the surface brightness profiles of the dwarf galaxies in our sample are generally too complicated to model with a simple Sérsic profile.

%Figure 9

\begin{figure}
    \centering
    \includegraphics[width=\linewidth]{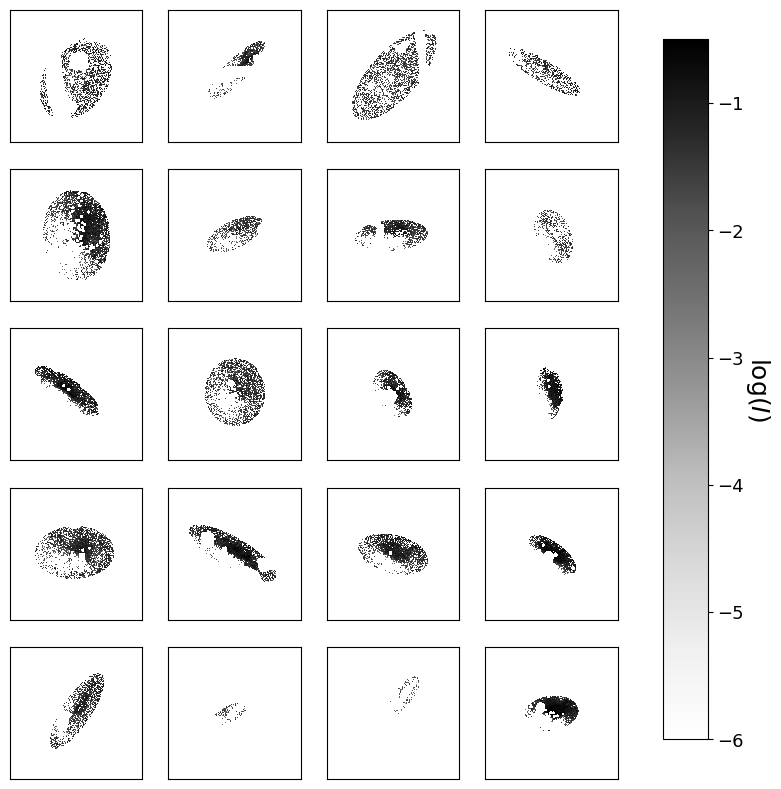}
    \caption{Normalized residual images created using the bicubic spline models. Each image has a $2r_e$ aperture applied around it. \textcolor{black}{Galaxies are ordered as in Fig.~\ref{NRI histogram figure}.}}
    \label{bicubic NRIs figure}
\end{figure}

\begin{figure}
    \centering
    \includegraphics[width=\linewidth]{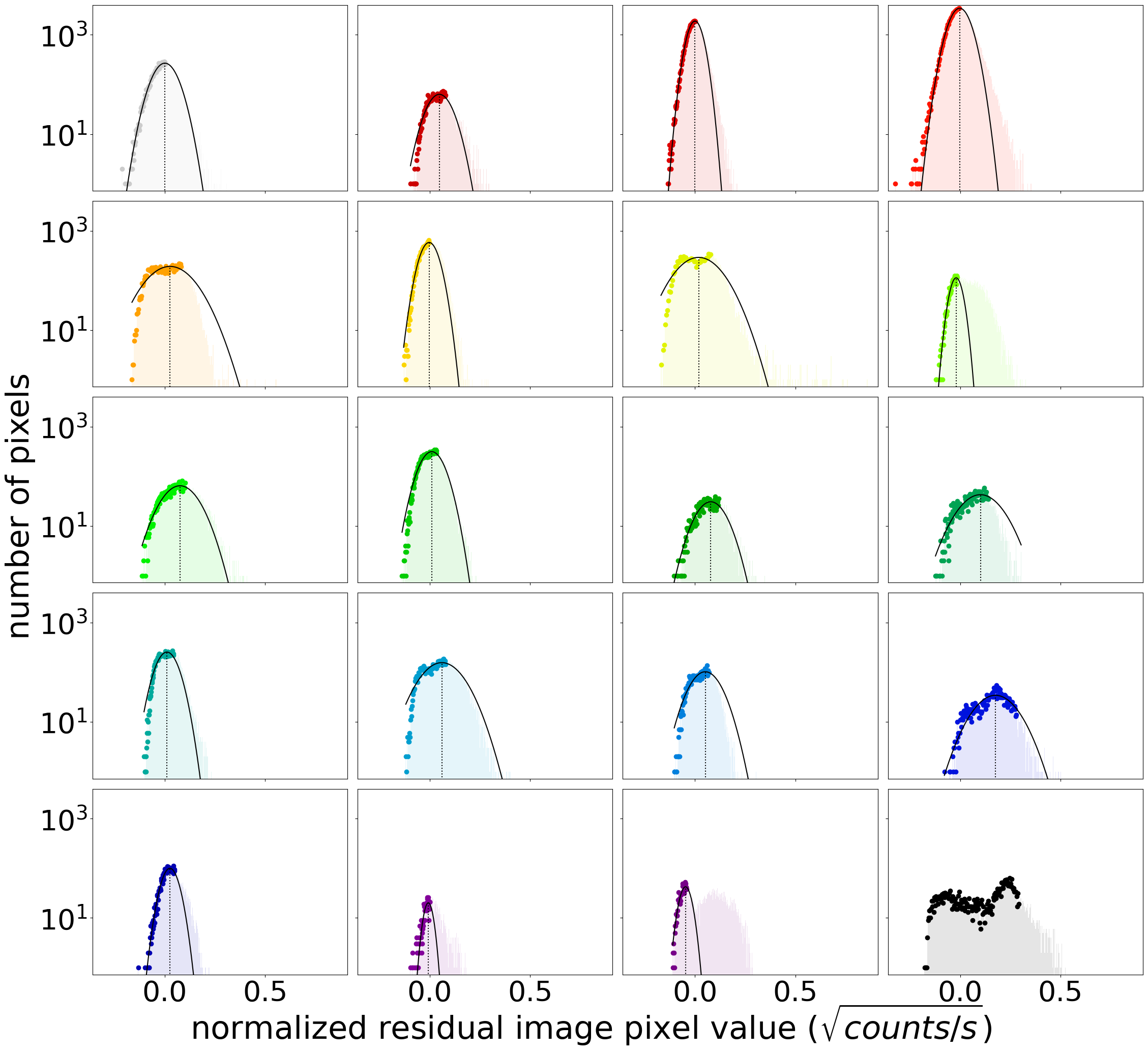}
    \caption{Pixel value distributions in the NRIs created using the bicubic spline model. \textcolor{black}{Galaxies are ordered as in Fig.~\ref{NRI histogram figure}.}}
    \label{bicubic NRI histogram figure}
\end{figure}

Like the \texttt{imfit} model, the bicubic spline model also leaves behind a lot of residual structure in the NRI (Fig.~\ref{bicubic NRIs figure}). We also found that many of the NRI pixel histograms created using the bicubic spline models (Fig.~\ref{bicubic NRI histogram figure}) do not show a clear {\color{black}Gaussian} component, and do not centre around 0 as expected. Furthermore, fitting the distribution for the furthest galaxy in the sample, 824852, did not produce a convergent result, and thus we could not obtain a SBF distance for this object. 

%Figure 10

%Figure 11

\begin{figure}
    \centering
    \includegraphics[width=\linewidth]{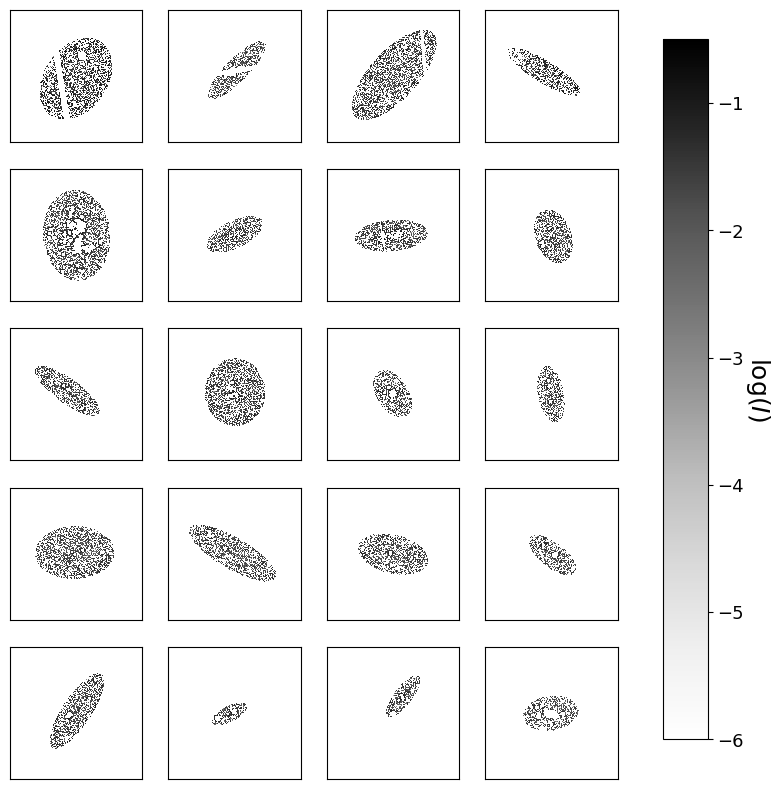}
    \caption{Normalized residual images created using the smoothed image models. Each image has a $2r_e$ aperture applied around it. \textcolor{black}{Galaxies are ordered as in Fig.~\ref{NRI histogram figure}.}}
    \label{smooth NRIs figure}
\end{figure}

By comparison, the smoothed-image model leaves much less residual structure in the NRI, as it is more customized to shape and gross features of the galaxy (Fig.~\ref{smooth NRIs figure}). In addition, the pixel value distributions of the NRIs created using the smoothed-image model (Fig.~\ref{NRI histogram figure}) show a {\color{black}Gaussian} component and centre around 0 as expected{\color{black}, assuming fluctuations are in the large-$N$ limit (see Section 3.1.6)}. Thus, we have adopted the smoothed-image model for our fiducial analysis, although we still use the Sérsic parameters given by \texttt{imfit} to position and scale the apertures around each galaxy.

%Figure 12

\subsection{Masking}
\label{mask tests section}
As discussed in Section \ref{masking and modelling section}, there are several ways to create a mask. Each represents a compromise between removing all contaminating sources and bad regions, and preserving genuine SBFs. Of our three masking options, the histogram mask is the most effective because it captures the most contaminating sources. 

\begin{figure}
    \centering
    \includegraphics[width=\linewidth]{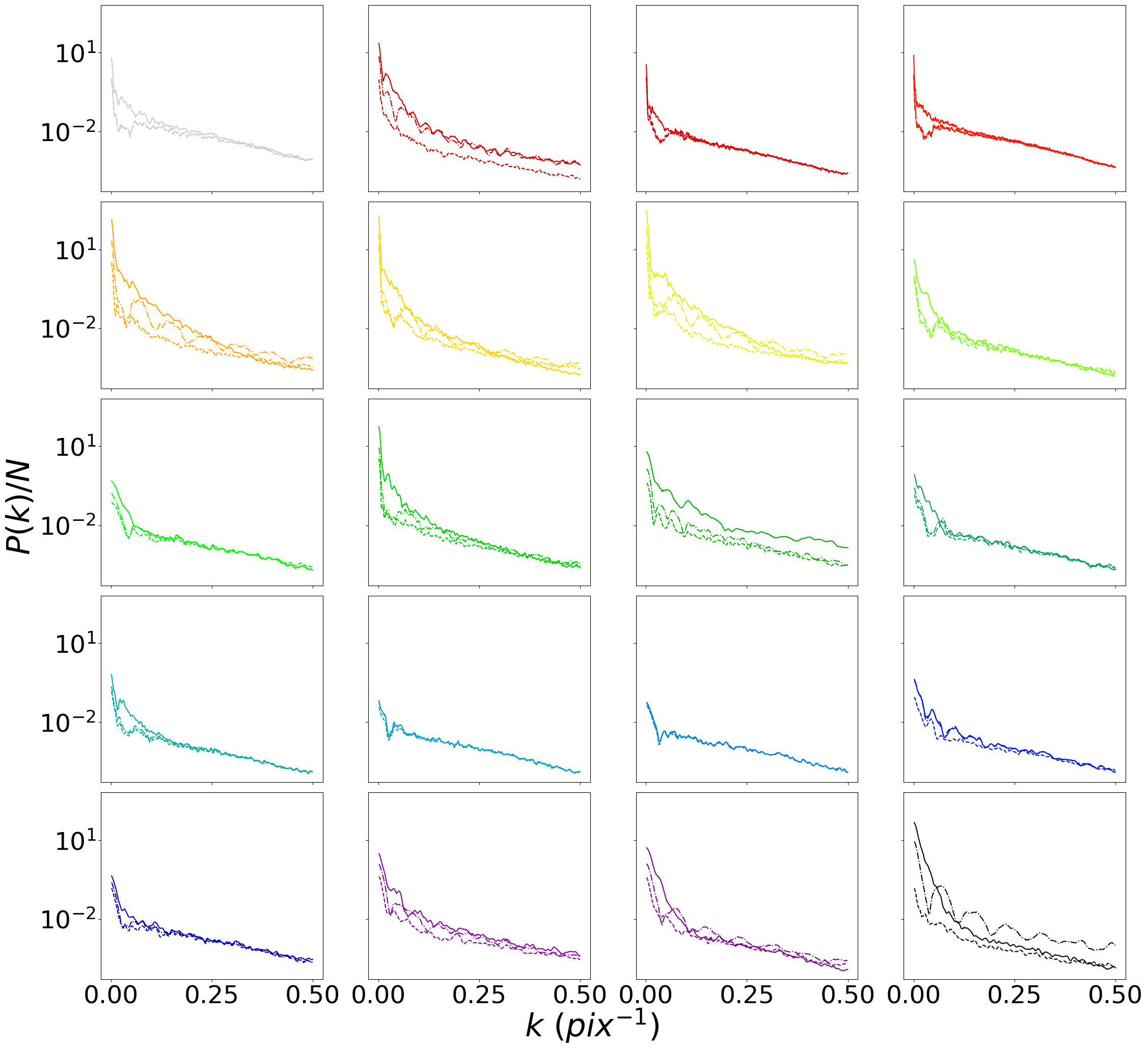}
    \caption{Power spectra of the masked normalized residual images using different masks. These are created using the smoothed image models, and a $1r_e$ aperture around the galaxy. Each subsequent masking method includes the previous method(s). \textcolor{black}{Galaxies are ordered as in Fig.~\ref{NRI histogram figure}.}}
    \label{masking options figure}
\end{figure}

The initial manual mask removes only the most obvious contaminants, while the automatic point source detection mask removes more point sources, but leaves noticeable oscillations in the power spectrum of the masked NRI (Fig.~\ref{masking options figure}). These appear to be particularly strong in objects with large numbers of bright central point sources. We have determined experimentally that the oscillations originate from the point-source masking itself. Using the same aperture size
for all the point sources detected introduces structure on this spatial scale, and thus produces periodic oscillations in the power spectrum, which will degrade the quality of the SBF distance estimate, as discussed further in appendix \ref{alernate masking appendix}. 

%Figure 13

To remove these features, an even more aggressive approach is required, so we introduce the histogram mask. This mask removes the oscillations seen in the automatic point-source detection mask spectra, and in some cases, covers all the region of the automatic mask, meaning a histogram masking approach could be effective on its own without further point source removal. Thus, we adopt the histogram mask (which includes our previous two masks) in our fiducial method.

\subsection{Background Subtraction}
\label{background tests section}
Examining our NRI power spectra, we found that the power does not become constant at large wavenumbers, but continues to decrease down to the Nyquist frequency. This is an artifact of our images, which have been combined from multiple pointings and exposures, correlating and smoothing pixel noise over a range of scales \citep{Mei2005, Mitzkus2018}. Thus, rather than fitting the NRI power spectrum using the original SBF equation (equation (\ref{SBF original equation})), we estimated the background from blank regions in the COSMOS field, as described above in Section \ref{power spectrum section}, and fit using equation (\ref{SBF background equation}).

To illustrate the differences between the two approaches, we created an average power spectrum by combining the individual NRI spectra of the 5 nearest galaxies, normalized to the same integrated power, and fit to the average using both background subtraction methods. In place of the expectation power spectrum, we used the power spectrum of the PSF alone, since the expectation power spectrum is only slightly different from the power spectrum of the PSF, and is different for each object since they have different masks. The fit was performed over the wavenumber range 0.1--0.3\,$\text{pix}^{-1}$. 

\begin{figure}
    \centering
    \includegraphics[width=\linewidth]{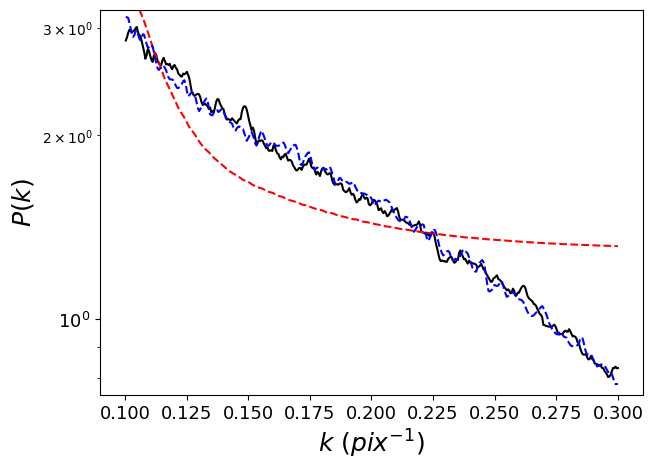}
    \caption{Average power spectrum of the masked normalized residual image for the 5 nearest objects (black curve), along with fits assuming a flat white-noise background (equation (\ref{SBF original equation}) -- red dashed curve), and a background determined directly from blank areas of the field (cf.~equation (\ref{SBF background equation}) -- blue dashed curve).}
    \label{representative fit figure}
\end{figure}

Fig.~\ref{representative fit figure} shows the the averaged masked NRI power spectrum (solid black curve), along with the two fits. It is clear that the fit using equation (\ref{SBF original equation}) (red dashed curve) does not accurately capture the shape of the power spectrum. Accounting for a realistic background contribution using equation (\ref{SBF background equation}), however (blue dashed curve), gives a very good fit to the spectrum, justifying its use in our fiducial method. \textcolor{black}{This approach is a significant improvement on the assumption of white noise, as it gives a much better fit to the data.}

%Figure 14

\subsection{Dependence on Aperture Size}
\label{aperture tests section}
The choice of aperture size to use when measuring SBFs is unclear a priori. Too large an aperture will dilute the signal with a larger background contribution, while too small an aperture may leave too little area to measure the SBFs. 

\begin{figure}
    \centering
    \includegraphics[width=\linewidth]{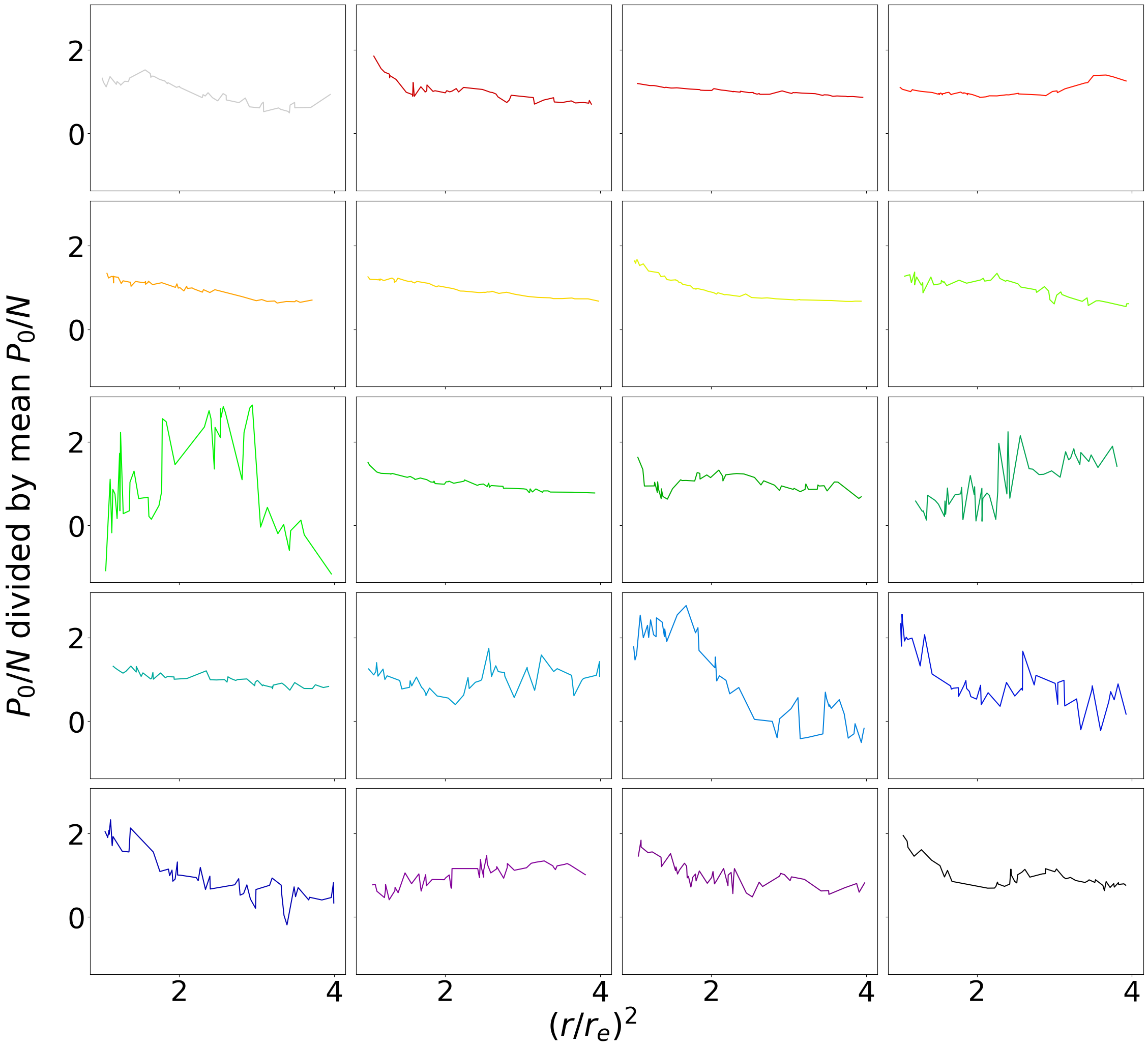}
    \caption{Dependence of $P_0/N$ on aperture size. The values of $P_0/N$ were computed using the smoothed-image model, with the histogram mask, and a fitting range $k = $ $0.1$ to $0.3\,\text{pix}^{-1}$. \textcolor{black}{Galaxies are ordered as in Fig.~\ref{NRI histogram figure}.}}
    \label{aperture dependence figure}
\end{figure}

As explained previously, to account for some of the uncertainty introduced by this choice, we run the SBF measurement process 50 times, choosing a random aperture size each time, taking the final value of $P_0/N$ to be the mean of the 50 resulting values, and its error to be the standard deviation of the distribution. Fig.~\ref{aperture dependence figure} shows the value of $P_0/N$ as a function of aperture size for each galaxy. In most cases, the variations seem small and/or random, but in some (e.g. 733922), we see indications of a systematic decrease in $P_0/N$ with aperture size. This could indicate excess power from point sources or smooth model errors in the centre of the image, and we have used these plots iteratively to test the effectiveness of our masking choices.

%Figure 15

\subsection{Wavenumber Range}
\label{wavenumber tests section}
As discussed in Section \ref{SBF variance section}, both the low and high wavenumber ends of the NRI power spectra are compromised by various factors, and thus need to be excluded from the fitting. We have tested many different wavenumber ranges, with lower bounds ranging from $0.08\,\text{pix}^{-1}$ to $0.20\,\text{pix}^{-1}$ and upper bounds ranging from $0.20\,\text{pix}^{-1}$ to $0.50\,\text{pix}^{-1}$. We note that using the smoothed image model removes some large-scale power from the NRI, introducing a dip in the power spectrum at values below $\sim0.10\,\text{pix}^{-1}$, so we need to set the lower end of the range to at least this value.

\begin{figure}
    \centering
    \includegraphics[width=\linewidth]{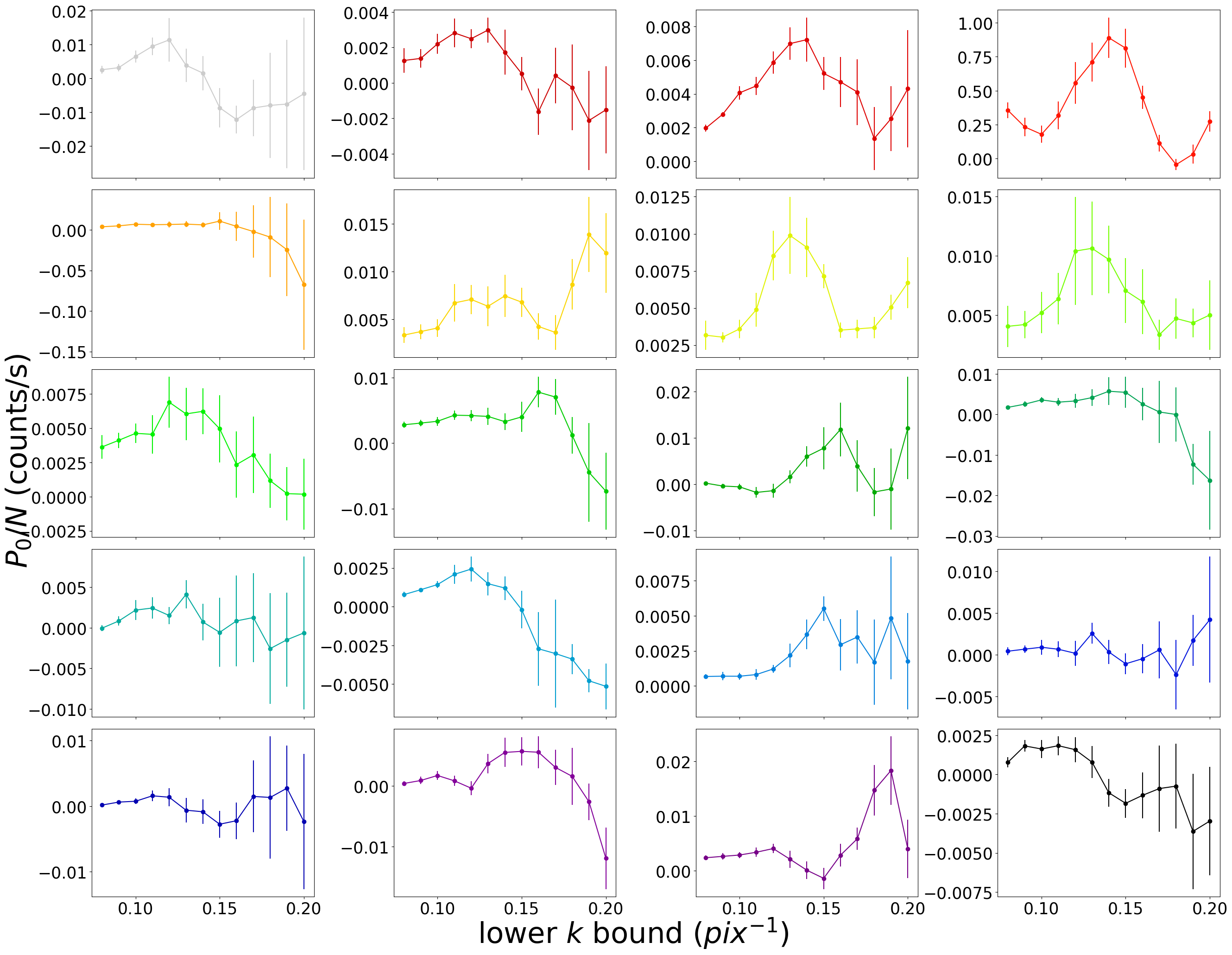}
    \caption{The normalized SBF signal plotted against the lower $k$ bound, given a fixed upper bound of $0.3\,\text{pix}^{-1}$. The error bars represent the standard deviation on the values obtained using 25 randomly selected apertures. \textcolor{black}{Galaxies are ordered as in Fig.~\ref{NRI histogram figure}.}}
    \label{lower bound figure}
\end{figure}

\begin{figure}
    \centering
    \includegraphics[width=\linewidth]{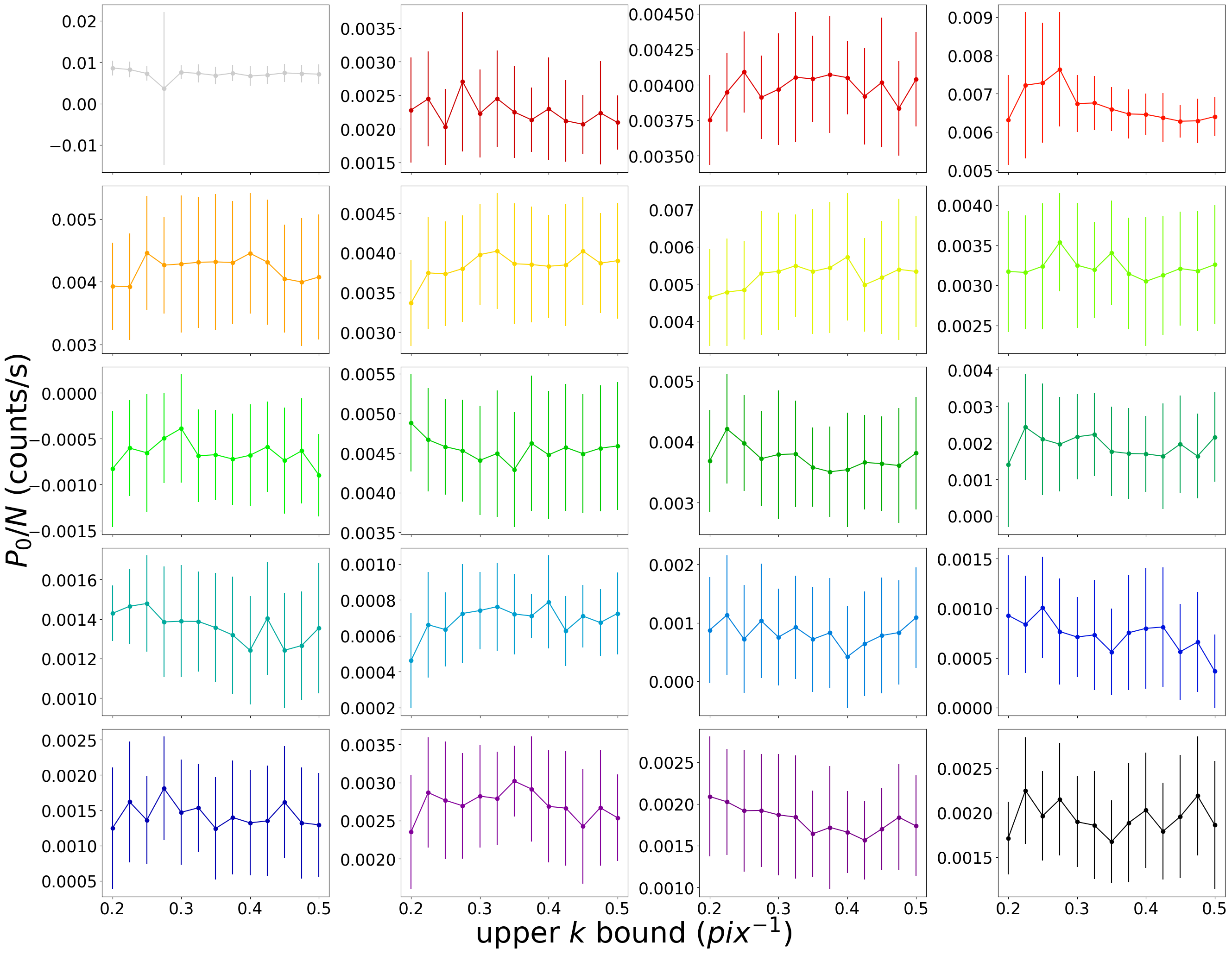}
    \caption{The normalized SBF signal plotted against the upper $k$ bound, given a fixed lower bound of $0.1\,\text{pix}^{-1}$. The error bars are as in Fig.~\ref{lower bound figure}. \textcolor{black}{Galaxies are ordered as in Fig.~\ref{NRI histogram figure}.}}
    \label{upper bound figure}
\end{figure}

To study the effect of wavenumber range, we plotted the values for $P_0/N$ against the lower bound, using a fixed upper bound of $0.30\,\text{pix}^{-1}$ in Fig.~\ref{lower bound figure}, and against the upper bound, using a fixed lower bound of $0.10\,\text{pix}^{-1}$ in Fig.~\ref{upper bound figure}. For each plot, for efficiency, one point represents 25 rather than 50 SBF runs. Otherwise the process is the same as described in Section \ref{methods section}. A flat curve in these plots means the SBF signal is stable against small changes in the range, which suggests it is a good region in which to select a bound. 

From Fig.~\ref{lower bound figure}, it is clear that the lower bound must be selected carefully as the SBF signal can vary significantly with the lower bound. We estimate that a lower bound in the region of $0.10$ to $0.13\,\text{pix}^{-1}$ is the best choice overall, as this region is stable across most of the galaxies. Fig.~\ref{upper bound figure} shows that the upper bound is much less important than the lower bound, as the plot is generally flat past about $0.30\,\text{pix}^{-1}$ for most galaxies. From this, we infer that the SBF signal dominates at low values of $k$, and that focusing on lower wavenumbers will give the best SBF fit. For the fiducial method, we choose to fit the wavenumber range $0.1$ to $0.3\,\text{pix}^{-1}$ to capture as much of the SBF signal as possible while excluding the dip in the power spectrum below $0.1\,\text{pix}^{-1}$.

%Figure 16

%Figure 17

\subsection{Colour Variation}
\label{colour tests section}

\begin{figure}
    \centering
    \includegraphics[width=\linewidth]{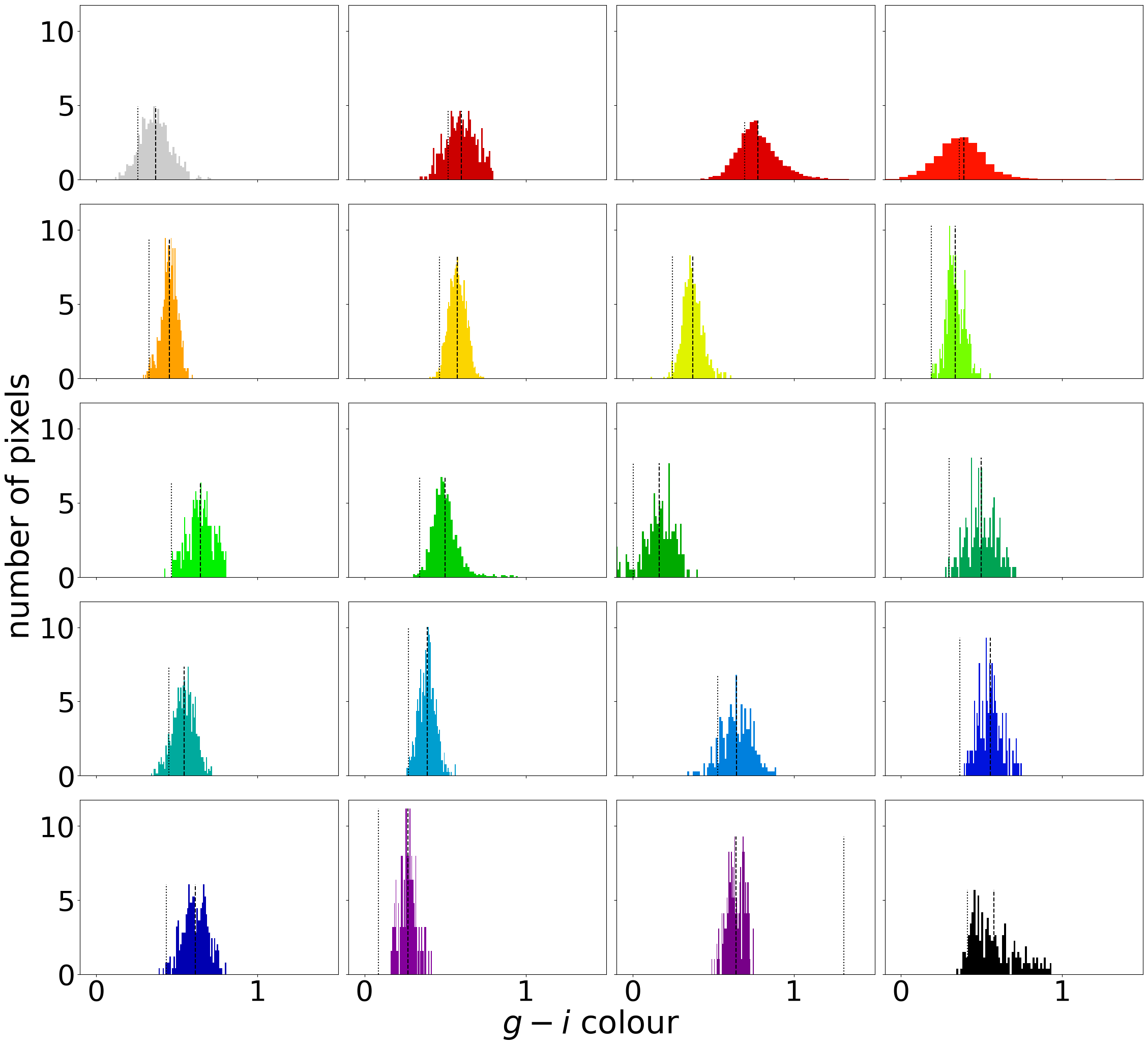}
    \caption{Pixel histograms for the $g-i$ images. The dotted vertical line represents the catalogue value and the dashed line represents the mean of the histogram values. The mean value and its corresponding standard deviation are what we choose to represent the $g-i$ value of the galaxy. \textcolor{black}{Galaxies are ordered as in Fig.~\ref{NRI histogram figure}.}}
    \label{g-i histogram figure}
\end{figure}

In Section \ref{SBF magnitude and distance section}, we outlined two methods for estimating galaxy colour. Both assume the same filter conversions, but one uses the catalogue magnitudes, while the other uses the images directly. For the second method, Fig.~\ref{g-i histogram figure} shows pixel histograms of the $g-i$ images within $1r_e$. It is clear from these histograms that in some cases, there is a large spread in the individual pixel colours relative to the mean catalogue colour, and that the latter may not correspond to the mean of the individual pixel colours. For these reasons, we choose to use the pixel-based approach in our fiducial method. The spread in some of the colour distributions (e.g. 824852) also suggests we should treat with caution SBF distances derived assuming a single mean colour\textcolor{black}{, as the colour can vary widely across the extent of the object. An improvement on the method might be to account for the changes in colour across the object, which could be explored in future work.}

%Figure 18

\section{Discussion}
\label{discussion section}

Dwarf galaxy populations in the local Universe represent a challenge and an opportunity to improve our understanding of galaxy formation, feedback, and dark matter physics. To identify local dwarfs and place them in their environmental context, distance estimates are essential, yet obtaining redshift-based distances for large numbers of faint, low-surface brightness dwarfs remains challenging. With a new generation of imaging surveys forthcoming, including space-based surveys with Euclid and Roman, it is worth reconsidering the potential for other methods such as SBF to tackle this problem. 

The SBF method has traditionally been applied to relatively bright, high surface-brightness early-type galaxies with relatively simple stellar populations \citep[][]{Moresco2022}, but in recent work, several groups have pushed to apply the technique to fainter systems within $\sim25$\,Mpc \citep[][]{Carlsten2019,  Polzin_2021, Kim2021, Greco_2021, Carlsten2022}. In this paper, we have tested the possibility of applying the SBF method to a set of 20 dwarf galaxies at distances of 20--150 Mpc, drawn from the HST/ACS imaging of the COSMOS survey. 
There are several important challenges in getting the method to work for this sample, including the overall faintness and low surface brightness of the galaxies, internal gradients or scatter in stellar populations and/or extinction, irregular morphology, and a correlated background component introduced by mosaicing. After making careful choices in how the SBF method is applied to the images, we find a {\color{black}Gaussian} SBF signal that does correlate linearly with distance, out to distances of 50--100 Mpc. On the other hand, the recovered SBF power is only $\sim\sfrac{1}{3}$ of that expected, leading to a corresponding overestimate in distance. 

Beyond 100 Mpc, the measured SBF power is generally too high, leading to underestimates in distance \textcolor{black}{(e.g., \cite{BlakesleePhD})}. Examining the point source luminosity functions of our most distant galaxies (Appendix \ref{luminosity function appendix}), we infer that this excess power comes from faint point sources undetected in the relatively shallow COSMOS imaging. \textcolor{black}{In principle, deeper observations that better resolve the point source luminosity function would overcome this limitation.}
 
Estimating SBF distances also requires a number of (only partly constrained) choices in the detailed method, including how to model the galaxy surface brightness profile, how to identify and mask contaminating sources, how to estimate and subtract the background contribution to the NRI power spectrum, the colour used in computing the SBF calibration magnitude, and the wavenumber range used in fitting the power spectrum. For the galaxies in our sample, many possible choices in these details lead to systematic variations significantly larger than the simplest errors we have estimated by varying the aperture around each galaxy slightly. Several choices increase the average SBF power for nearby galaxies until it is closer to the expected level, but in doing so produce {\color{black} oscillations in the NRI power spectrum, non-Gaussian features in the NRI pixel distribution}, larger uncertainties, and/or 
reduce the correlation between detected power and actual distance.

Recent work by \citet{Polzin_2021} estimated a SBF distance to the nearest galaxy in our sample, 549719. They obtained a result in agreement with the proper distance ($d_c=21.3\text{Mpc}$), $d_{\rm SBF}=24\pm3\text{Mpc}$ (their full galaxy measurement, which averages over two regions with slightly different colours). By comparison, with our fiducial method, we obtain a SBF distance of $d_{\rm SBF}=56\text{Mpc}$ with lower and upper limits of $50\text{Mpc}$ and $67\text{Mpc}$ respectively. Thus, our fiducial estimate is incorrect by a factor of almost 3, and inconsistent with the \citet{Polzin_2021} result. We can recover their result by modifying our methodology to make it more similar to theirs. If we use a single Sérsic profile to model the smooth light distribution, flat background subtraction assuming the standard SBF equation (Eq.~\ref{SBF original equation}), a less aggressive mask (using only a manual mask modelled after their figure 2), and a wavenumber range of $0.15$ to $0.4\,\text{pix}^{-1}$, we obtain a SBF distance of $23\pm5\,\text{Mpc}$, in excellent agreement with their result.
These are not optimal choices for the rest of our sample, however; if we apply this modified method to the whole sample, we find all our galaxies have estimated SBF distances in the range of $\sim$10--30\,Mpc, and there is little or no correlation between $d_{\rm SBF}$ and $d_c$.

There are a few ways we may imagine making progress on the problem of generalizing the SBF method to a more diverse, distant, low surface-brightness populations of galaxies. With larger samples of partially resolved objects, we might be able to develop a new calibration for our current fiducial method, that corrected for the zero-point offset seen in Fig.~\ref{dsbf vs dz figure}. We note that with a different calibration, most objects in this figure with true distances of less than 100 Mpc might have SBF distances consistent to within the (10-15\%) statistical errors shown in this plot.

We also have several important pieces of additional information about the galaxies in our sample, including their associated point source luminosity functions (see Appendix \ref{luminosity function appendix}), their SEDs (see Appendix \ref{galaxy property appendix}), and the detailed shape of the pixel histogram (Fig.~\ref{NRI histogram figure}). These diagnostic features may allow corrections to the SBF method for age, metallicity, galaxy type and/or star formation history. More general statistical methods that incorporate the additional information, such as those used in machine learning (ML), might be be applied to this problem, in order to determine the best choices to make in applying the SBF method, and which factors are most important in the analysis \citep[see][for recent examples of the application of ML to the morphological classification of low surface brightness dwarf galaxies]{Tanoglidis2020, Mueller2021}. Such methods might also be used to identify and correct the NRI in distant galaxies where undetected point sources contribute significant power, as discussed in Appendix \ref{luminosity function appendix}. We will explore these possibilities in future work.

Given the imaging data expected from forthcoming surveys \citep{Moresco2022}, extending the SBF method to a broader class of galaxies is a priority. The Euclid Wide Survey \citep{Euclid}, for example, will cover 15,000 square degrees on the sky, almost $10^4$ times the effective area of the COSMOS ACS imaging, albeit with lower resolution ($\sim$ 0.2'') and less collecting area. If the 20--30 galaxies detected in the COSMOS field at distances of less then 200 Mpc are indicative, Euclid may see thousands of partially-resolved dwarf galaxies out to $\sim$50 Mpc. The Roman Space Telescope \citep{Roman}, with 100 times the field of view of HST and better infrared sensitivity, could produce samples of thousands of objects out to distances of 200 Mpc or beyond. From the ground, the Vera Rubin Observatory \citep{Rubin} will similarly produce large samples of partially resolved galaxies over the distance range of the COSMOS sample \citep{Greco_2021}. These samples could provide a trove of information about galaxy formation on the smallest scales, if we can identify these nearby objects and correctly place them in their three-dimensional context.

\section*{Acknowledgements}

We thank M. Hudson\textcolor{black}{, L. Parker, M. Bravo, D. Lazarus, and M. Oxland for useful discussions. We also thank the anonymous referee for comments which significantly improved this work.} JET acknowledges support from the Natural Sciences and Engineering Research Council (NSERC) of Canada through a Discovery Grant. The COSMOS 2015 catalogue is based on data products from observations made with European Southern Observatory (ESO) Telescopes at the La Silla Paranal Observatory under ESO programme ID 179. A-2005, and on data products produced by TERAPIX and the Cambridge Astronomy Survey Unit on behalf of the UltraVISTA consortium.

This research made use of the NASA/IPAC Infrared Science Archive and the NASA/IPAC Extragalactic Database (NED), which are funded by the National Aeronautics and Space Administration and operated by the California Institute of Technology, as well as the following software: \texttt{astropy} \citep{astropy:2013,astropy:2018}, \texttt{DS9} \citep{ds9}, \texttt{imfit} \citep{Erwin_2015}, \texttt{matplotlib} \citep{matplotlib}, \texttt{numpy} \citep{numpy}, \texttt{photutils} \citep{larry_bradley_2022_6825092}, and \texttt{scipy} \citep{scipy}.

%%%%%%%%%%%%%%%%%%%%%%%%%%%%%%%%%%%%%%%%%%%%%%%%%%

\section*{Data Availability}

The data used in this article are publicly available. The COSMOS 2015 catalogue \citep{Laigle_2016} can be accessed from the COSMOS website, at  \url{http://cosmos.astro.caltech.edu/page/photom}. Spectroscopic redshifts for our sample were taken from \citet{Xi2018}, and are generally also available in NED. The derived data generated in this work will also be shared on reasonable request to the corresponding author.


\bibliographystyle{mnras}
\bibliography{mybib}

%%%%%%%%%%%%%%%%%%%%%%%%%%%%%%%%%%%%%%%%%%%%%%%%%%

%%%%%%%%%%%%%%%%% APPENDICES %%%%%%%%%%%%%%%%%%%%%

\appendix

\section{Properties of the Galaxy Sample} % should this go in the main text?
\label{galaxy property appendix}

Our sample was drawn from the COSMOS survey, a multi-wavelength survey including space-based ACS imaging and deep ground-based imaging. We used the COSMOS2015 catalogue \citep{Laigle_2016} for integrated magnitudes and other properties of the sample, although we note that more recent photometry has been published by \citet{Weaver2022}.

\subsection{Imaging and photometric properties}

The mosaic imaging used in the SBF analysis was obtained from the NASA/IPAC IRSA COSMOS Cutouts Service in the HST/ACS F814W band \citep{Koekemoer_2007, Massey_2010}, and the Subaru $i^+$, IA464, and IA484 bands \citep{Taniguchi_2015}. The HST imaging is used for the SBF measurement itself, while the Subaru imaging is used to estimate the $g-i$ colour of the galaxies. The F814W imaging has units of counts/s, a pixel scale of 0.03$\arcsec$/pixel, and a zeropoint of 25.94. The zeropoint was calculated using equation (\ref{zeropoint equation}):
\begin{equation}
    ZP = -2.5\log_{10}(PHOTFLAM) - 5\log_{10}(PHOTPLAM) - 2.408
    \label{zeropoint equation}
\end{equation}
from \citep{stsci}, and the corresponding values from the image headers).
The Subaru imaging has units of counts, a pixel scale of 0.15''/pixel, and a zeropoint of 31.4 \citep{subaru_info}.

The \texttt{imfit} model parameters for the individual galaxies are given in Table \ref{property table}. The apparent magnitude in the F814W band was computed using
\begin{equation}
    m = m_{\rm zpt} -2.5\log_{10}{\left( 2\pi I_er_e^2n\frac{\exp{(b)}}{b^{2n}}\Gamma(2n)(1-e)\,, \right)}
    \label{sersic magnitude equation}
\end{equation}
where $b$ is computed using the approximation
\begin{equation}
    b = 2n - \frac{1}{3} + \frac{4}{405n} + \frac{46}{25515n^2}
    \label{b approximation equation}
\end{equation}
\citep{ciotti}. Table \ref{property table} also lists the absolute magnitude, calculated using the model apparent magnitude and the proper distance, and the stellar mass as given by the COSMOS2015 catalogue.

\begin{table*}
    \centering
    \caption{Galaxy properties. Position angle $\theta$ is given in degrees counter clockwise from the positive y-axis.}
    \begin{tabular}{c | c | c | c | c | c | c | c | c}

    COSMOS ID & $I_e$ (counts/s) & $r_e$ (pix) & $n$ & $e$ & $\theta$ (degrees) & $m_{F814W}$ & $M_{F814W}$ & $M_*/M_\odot$ \\\hline
    424575 & $0.02570\pm0.00008$ & $123.7\pm0.2$ & $0.815\pm0.004$ & $0.7089\pm0.0006$ & $315.30\pm0.04$ & 18.19 & -13.80 & $(3.4\pm0.3)\times10^8$ \\
    549719 & $0.00453\pm0.00003$ & $100.8\pm0.5$ & $0.387\pm0.006$ & $0.290\pm0.006$   & $324.9\pm0.7$   & 19.86 & -12.16 & $(7.8\pm0.7)\times10^7$ \\
    677414 & $0.0057\pm0.0001$   & $286\pm2$     & $0.74\pm0.02$   & $0.576\pm0.003$   & $137.45\pm0.08$ & 17.65 & -14.65 & $(8.0\pm0.7)\times10^8$ \\
    260583 & $0.00450\pm0.00001$ & $422.6\pm0.6$ & $0.443\pm0.002$ & $0.7323\pm0.0008$ & $58.82\pm0.05$  & 17.76 & -14.60 & $(2.2\pm0.2)\times10^8$ \\
    561851 & $0.02195\pm0.00006$ & $85.9\pm0.1$  & $0.706\pm0.003$ & $0.261\pm0.001$   & $4.7\pm0.2$     & 18.21 & -14.19 & $(2.0\pm0.2)\times10^8$ \\ 
    401988 & $0.02149\pm0.00002$ & $174.0\pm0.1$ & $0.700\pm0.001$ & $0.5575\pm0.0005$ & $116.05\pm0.03$ & 17.26 & -15.18 & $(7\pm1)\times10^8$ \\
    733922 & $0.02479\pm0.00006$ & $138.2\pm0.2$ & $0.874\pm0.003$ & $0.5726\pm0.0006$ & $95.14\pm0.05$  & 17.54 & -15.07 & N/A \\
    686606 & $0.0110\pm0.0005$   & $70\pm2$      & $0.66\pm0.05$   & $0.35\pm0.02$     & $19.9\pm0.6$    & 19.58 & -13.07 & $(2.0\pm0.2)\times10^7$ \\
    213165 & $0.01040\pm0.00003$ & $105.8\pm0.2$ & $0.857\pm0.003$ & $0.102\pm0.002$   & $0\pm0$         & 18.27 & -15.10 & $(2.2\pm0.2)\times10^8$ \\
    259971 & $0.0086\pm0.0005$   & $74\pm1$      & $0.68\pm0.07$   & $0.681\pm0.005$   & $54.1\pm0.2$    & 20.49 & -12.90 & $(4.4\pm0.4)\times10^7$ \\
    709026 & $0.00791\pm0.00007$ & $49.5\pm0.4$  & $1.00\pm0.01$   & $0.396\pm0.004$   & $32.2\pm0.4$    & 20.58 & -13.15 & $(1.3\pm0.1)\times10^7$ \\
    918161 & $0.01130\pm0.00009$ & $53.7\pm0.3$  & $0.617\pm0.008$ & $0.545\pm0.003$   & $10.5\pm0.2$    & 20.53 & -13.23 & $(2.7\pm0.3)\times10^7$ \\
    458976 & $0.00974\pm0.00003$ & $99.9\pm0.2$  & $0.718\pm0.002$ & $0.334\pm0.002$   & $91.9\pm0.2$    & 18.87 & -15.09 & $(1.9\pm0.2)\times10^8$ \\
    331749 & $0.00908\pm0.00004$ & $125.2\pm0.4$ & $0.643\pm0.005$ & $0.684\pm0.001$   & $59.78\pm0.08$  & 19.31 & -15.38 & $(8.0\pm0.7)\times10^7$ \\
    880547 & $0.00572\pm0.00003$ & $67.3\pm0.3$  & $0.686\pm0.005$ & $0.457\pm0.003$   & $76.1\pm0.3$    & 20.55 & -14.63 & $(5.3\pm0.7)\times10^7$ \\
    380820 & $0.01451\pm0.00008$ & $53.9\pm0.2$  & $0.654\pm0.005$ & $0.564\pm0.002$   & $52.1\pm0.1$    & 20.28 & -14.95 & $(3.6\pm0.3)\times10^7$ \\
    589205 & $0.00822\pm0.00004$ & $83.3\pm0.3$  & $0.684\pm0.005$ & $0.672\pm0.002$   & $326.4\pm0.1$   & 20.24 & -15.00 & $(4.4\pm0.8)\times10^7$ \\
    279307 & $0.01114\pm0.00004$ & $49.0\pm0.1$  & $0.041\pm0$     & $0.5918\pm0.0006$ & $297.0\pm0.1$   & 21.03 & -14.27 & $(1.9\pm0.2)\times10^7$ \\
    377112 & $0.01739\pm0.00009$ & $63.5\pm0.2$  & $0.848\pm0.006$ & $0.642\pm0.002$   & $322.50\pm0.08$ & 19.82 & -15.60 & $(1.12\pm0.09)\times10^8$ \\
    824852 & $0.0331\pm0.0004$   & $52.5\pm0.2$  & $1.09\pm0.01$   & $0.375\pm0.004$   & $278.5\pm0.2$   & 18.82 & -16.75 & $(1.8\pm0.1)\times10^8$ \\
    
    \end{tabular}
    \label{property table}
\end{table*}

\setlength{\tabcolsep}{10pt} % Default value: 6pt
\renewcommand{\arraystretch}{1.5} % Default value: 1
\begin{table*}
    \color{black}
    \centering
    \caption{SBF results. SBF magnitudes are given in the F814W band.}
    \begin{tabular}{c | c | c | c | c}

    COSMOS ID & $g-i$ & $M_{\text{SBF}}$ & $m_{\text{SBF}}$ & $d_{\text{SBF}}$ \\\hline

    424575 & $ 0.6 \pm 0.093 $ & $ -1.9 \pm 0.3 $ & $ 32.6 ^{+ 0.3 }_{- 0.2 }$ & $ 81.4 ^{+ 11.8 }_{- 8.3 }$ \\
    549719 & $ 0.367 \pm 0.094 $ & $ -2.4 \pm 0.3 $ & $ 31.3 ^{+ 0.3 }_{- 0.3 }$ & $ 56.3 ^{+ 10.4 }_{- 6.7 }$ \\
    677414 & $ 0.777 \pm 0.128 $ & $ -1.5 \pm 0.4 $ & $ 31.9 ^{+ 0.1 }_{- 0.0 }$ & $ 51.7 ^{+ 2.1 }_{- 1.8 }$ \\
    260583 & $ 0.39 \pm 0.21 $ & $ -2.3 \pm 0.5 $ & $ 31.4 ^{+ 0.1 }_{- 0.1 }$ & $ 57.8 ^{+ 3.9 }_{- 3.3 }$ \\
    561851 & $ 0.454 \pm 0.049 $ & $ -2.2 \pm 0.3 $ & $ 31.8 ^{+ 0.2 }_{- 0.2 }$ & $ 65.2 ^{+ 7.5 }_{- 5.6 }$ \\
    401988 & $ 0.575 \pm 0.052 $ & $ -1.9 \pm 0.3 $ & $ 31.9 ^{+ 0.2 }_{- 0.2 }$ & $ 60.9 ^{+ 6.2 }_{- 4.7 }$ \\
    733922 & $ 0.372 \pm 0.061 $ & $ -2.4 \pm 0.3 $ & $ 31.5 ^{+ 0.4 }_{- 0.2 }$ & $ 63.6 ^{+ 12.3 }_{- 7.8 }$ \\
    686606 & $ 0.337 \pm 0.056 $ & $ -2.4 \pm 0.3 $ & $ 32.1 ^{+ 0.3 }_{- 0.2 }$ & $ 85.7 ^{+ 11.9 }_{- 8.4 }$ \\
    213165 & $ 0.499 \pm 0.08 $ & $ -2.1 \pm 0.3 $ & $ 31.8 ^{+ 0.2 }_{- 0.2 }$ & $ 62.0 ^{+ 5.9 }_{- 4.6 }$ \\
    259971 & $ 0.645 \pm 0.081 $ & $ -1.8 \pm 0.3 $ & $ 36.1 + \infty $ & $ 386.1 + \infty$ \\
    709026 & $ 0.165 \pm 0.098 $ & $ -2.8 \pm 0.3 $ & $ 32.0 ^{+ 0.2 }_{- 0.2 }$ & $ 99.3 ^{+ 11.7 }_{- 8.7 }$ \\
    918161 & $ 0.499 \pm 0.092 $ & $ -2.1 \pm 0.3 $ & $ 32.7 ^{+ 0.9 }_{- 0.5 }$ & $ 92.8 ^{+ 51.1 }_{- 19.1 }$ \\
    458976 & $ 0.543 \pm 0.067 $ & $ -2.0 \pm 0.3 $ & $ 33.1 ^{+ 0.2 }_{- 0.1 }$ & $ 109.9 ^{+ 9.4 }_{- 7.5 }$ \\
    331749 & $ 0.389 \pm 0.051 $ & $ -2.3 \pm 0.3 $ & $ 33.7 ^{+ 0.4 }_{- 0.2 }$ & $ 169.9 ^{+ 29.5 }_{- 19.4 }$ \\
    880547 & $ 0.645 \pm 0.095 $ & $ -1.8 \pm 0.4 $ & $ 33.6 ^{+ \infty }_{- 0.7 }$ & $ 126.1 ^{+ \infty }_{- 37.7 }$ \\
    380820 & $ 0.553 \pm 0.074 $ & $ -2.0 \pm 0.3 $ & $ 33.6 ^{+ 1.1 }_{- 0.6 }$ & $ 134.5 ^{+ 91.1 }_{- 29.6 }$ \\
    589205 & $ 0.614 \pm 0.078 $ & $ -1.8 \pm 0.3 $ & $ 33.1 ^{+ 0.9 }_{- 0.5 }$ & $ 101.2 ^{+ 55.6 }_{- 20.8 }$ \\
    279307 & $ 0.269 \pm 0.052 $ & $ -2.6 \pm 0.2 $ & $ 32.3 ^{+ 0.3 }_{- 0.2 }$ & $ 101.1 ^{+ 16.4 }_{- 11.0 }$ \\
    377112 & $ 0.639 \pm 0.055 $ & $ -1.8 \pm 0.3 $ & $ 32.8 ^{+ 0.4 }_{- 0.3 }$ & $ 85.2 ^{+ 16.9 }_{- 10.6 }$ \\
    824852 & $ 0.576 \pm 0.138 $ & $ -1.9 \pm 0.4 $ & $ 32.8 ^{+ 0.4 }_{- 0.3 }$ & $ 90.2 ^{+ 18.0 }_{- 11.3 }$ \\
    
    \end{tabular}
    \label{SBF table}
\end{table*}

\subsection{Spectral Energy Distributions}
\begin{figure}
    \centering
    \includegraphics[width=\linewidth]{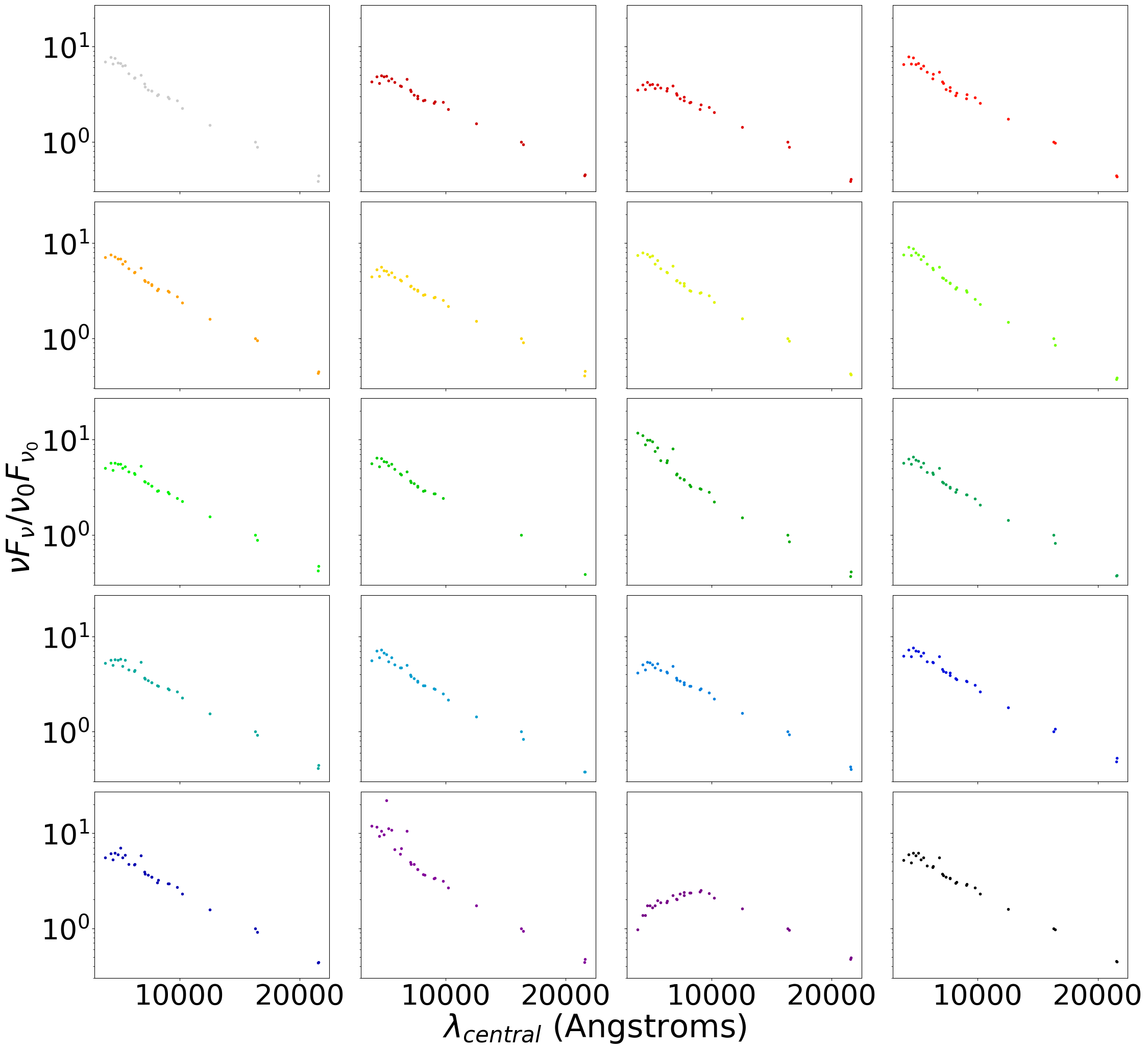}
    \caption{SEDs for each galaxy in the sample, normalized to the value at $16311$Å. \textcolor{black}{Galaxies are ordered as in Fig.~\ref{NRI histogram figure}.}}
    \label{all SEDs figure}
\end{figure}

%\label{SED appendix}
To test whether any of the systematics in our derived distance estimates might be related to galaxy type, 
we measured spectral energy distributions (SEDs) for each galaxy in the sample, using 3 arcsecond aperture magnitudes from the COSMOS 2015 catalogue \citep{Laigle_2016}. The SEDs, normalized to the value at $x=16311$Å, are shown in Fig.~\ref{all SEDs figure}. Horizontal error bars represent the width of the filter, while vertical error bars indicate errors on the photometry. We see broadly similar SED types in most cases, but also a few outliers, such as the relatively young/star-forming object 279307, or the relatively passive/old 377122. Unfortunately, comparing SED type to residuals in the $d_{\rm SBF}$--$d_{c}$ relation did not reveal any clear trends, although it might do so with a larger sample. 

%Figure A1

\section{Point Source Luminosity Functions}
\label{luminosity function appendix}
\begin{figure*}
    \centering
    \includegraphics[width=\linewidth]{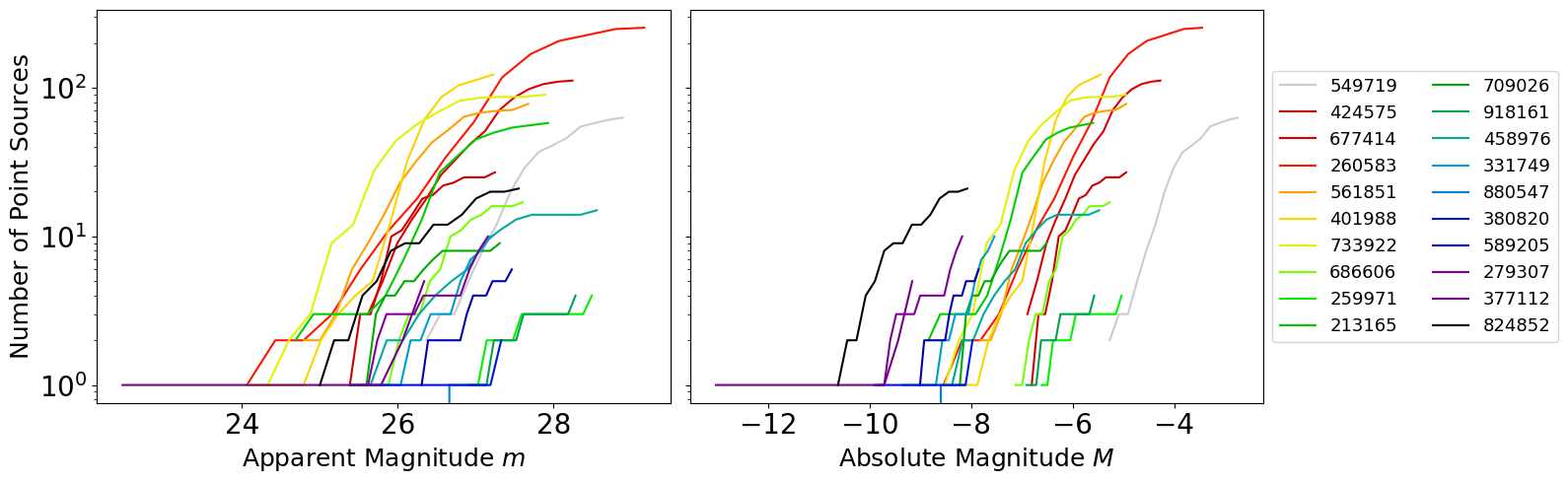}
    \caption{Luminosity functions for each galaxy in the sample. Left: in terms of apparent magnitude. Right: in terms of absolute magnitude, calculated using the proper distance.}
    \label{luminosity functions figure}
\end{figure*}

In addition to SBFs, most of the galaxies in our sample also contain populations of bright point sources, which may be globular clusters, unresolved HII regions, young star clusters, or even individual young stars. These sources contribute additional power to the NRI, unrelated to SBFs, so it is important to identify and mask them. We detected sources using \texttt{photutils}, as described in Section \ref{masking and modelling section}. The resulting luminosity functions are shown in Fig.~\ref{luminosity functions figure}, both in terms of apparent magnitude (left-hand panel), and in terms of absolute magnitude (right-hand panel, using a distance modulus based on the proper distance). The effective depth of the luminosity functions varies due to differences in the background and smooth galaxy model, but is generally in the range $m_{814W} = 26$--27.5, judging from the slope of the counts. For the nearer galaxies in our sample, this corresponds to absolute magnitudes $-6$ to $-4$, around the bottom of the globular cluster luminosity function, and the beginning of the red giant branch (RGB) whose stars contribute strongly to SBF. \cite{Carlsten2019} have argued that masking to this depth should eliminate essentially all globular clusters expected in dwarf galaxies. On the other hand, \cite{Kim2021} recently tested the effect of point source masking to greater depths on a sample of nearby dwarfs, observed with Hyper Suprime-Cam. Masking all point sources down to a depth of $M_g = -3.5$ to $-5$, they still find significant variations in the SBF absolute magnitude over this range, at least for the bluer part of their sample ($g-i < 0.5$). Thus, for the nearest galaxies in our sample, it seems we have just enough depth to account for point source contamination, and may fall a bit short of this for blue galaxies. 

On the other hand, for the most distant galaxies located at 100 Mpc or more, the point source depth of the COSMOS imaging corresponds to  absolute magnitudes $-10$ to $-8$, suggesting these are bright globular clusters or HII regions, and many fainter point sources remain undetected. We note that the 3--4 most distant galaxies are also those for which we measure significantly more power in the NRI (see Fig.~\ref{Power_vs_expected}). It seems likely that this power ("excess", relative to the power measured in the rest of the sample, though the resulting total power is close to the SBF expectation) comes from undetected fainter point sources in these objects.  

While we cannot detect and mask out these fainter sources without deeper data, close examination of the apparent luminosity functions suggests we may be able to identify galaxies where undetected point sources are a problem. For the nearest galaxies in the sample, we detect many point sources, and thus can remove much of the associated power from the NRI with masking. For some more distant galaxies, the only detected point sources are very faint, and thus any undetected sources will contribute relatively little power. The difficult cases seem to be those where we detect 5--10 fairly bright point sources, but few fainter ones (e.g.~galaxies 824852, 37712, 279307, and 589205). This suggests that with further consideration of the shape of the point source luminosity function, together with colours, radial distributions, and/or general information about the galaxy (magnitude, surface brightness, concentration, etc.), we might be able to distinguish nearby galaxies from more distant ones, e.g.~using a machine-learning approach trained on a larger sample, and correct for undetected point sources statistically \citep[e.g.][]{Tonry_1988,Mei2005}.

%Figure B1

\section{Effect of Alternate Masking Strategies}
\label{alernate masking appendix}

\begin{figure}
    \centering
    \includegraphics[width=\linewidth]{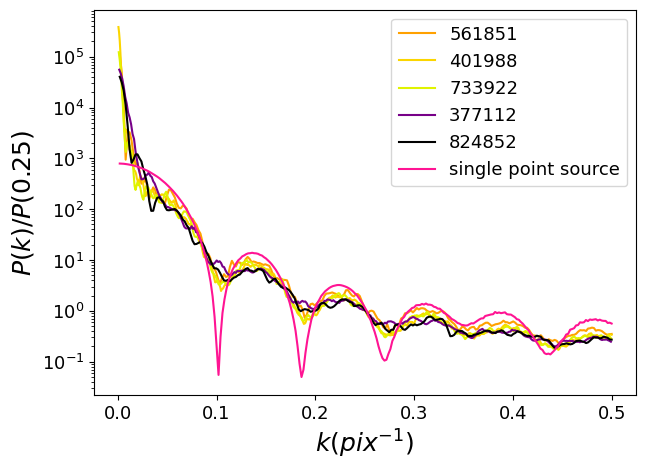}
    \caption{Power spectrum of the automatic point source mask, for each of the subset of the sample with the strongest oscillations. The power spectrum of a single point-source mask (solid pink curve) is shown for comparison.}
    \label{mask power spectra oscillations figure}
\end{figure}

To determine the origin of the oscillations in the masked NRI power spectra, we created a mask for a single automatically-detected point source (one circle with a 6 pixel radius). We then compared the power spectrum of this single-source mask with the power spectrum of their full mask, as well as the power spectrum of their masked NRI (Fig.~\ref{mask power spectra oscillations figure}). We see clearly that the oscillations in the power spectrum of the single-source mask correspond closely to the oscillations in the other power spectra. Thus, we conclude that these oscillations originate from the automatic point source mask. While our method includes this point source mask in the expectation power spectrum, we suspect that noise in the power spectrum around the oscillations leads to poor fitted values for the automatic point-source mask (cf.~\ref{distance results for alternate masking figure}). 

%Figure C1
    
%Figure C2
\begin{figure}
    \centering
    \includegraphics[width=\linewidth]{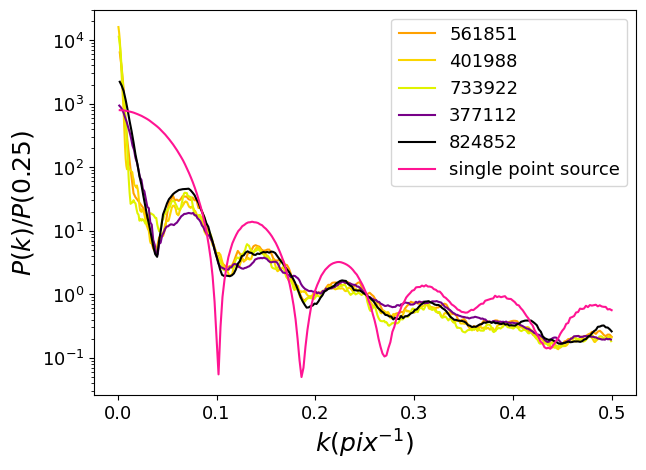}
    \caption{Power spectrum of the NRI, for each of the subset of the sample with the strongest oscillations. The power spectrum of a single point-source mask (solid pink curve) is shown for comparison.}
    \label{nri power spectra oscillations figure}
\end{figure}

We tested two alternate versions of the automatic mask, to try to remove the oscillations in the spectra. First, we simply smoothed the automatic mask using the same Gaussian kernel used for the smoothed image model (see section \ref{smoothed image model section}). This was not effective, as it did not fully mask the point sources, i.e. the masked areas around these objects were no longer set to 0. A second method was to multiply this smoothed mask by the original mask, effectively smoothing the edges of each masked region.

\begin{figure*}
    \centering
    \includegraphics[width=0.8\linewidth]{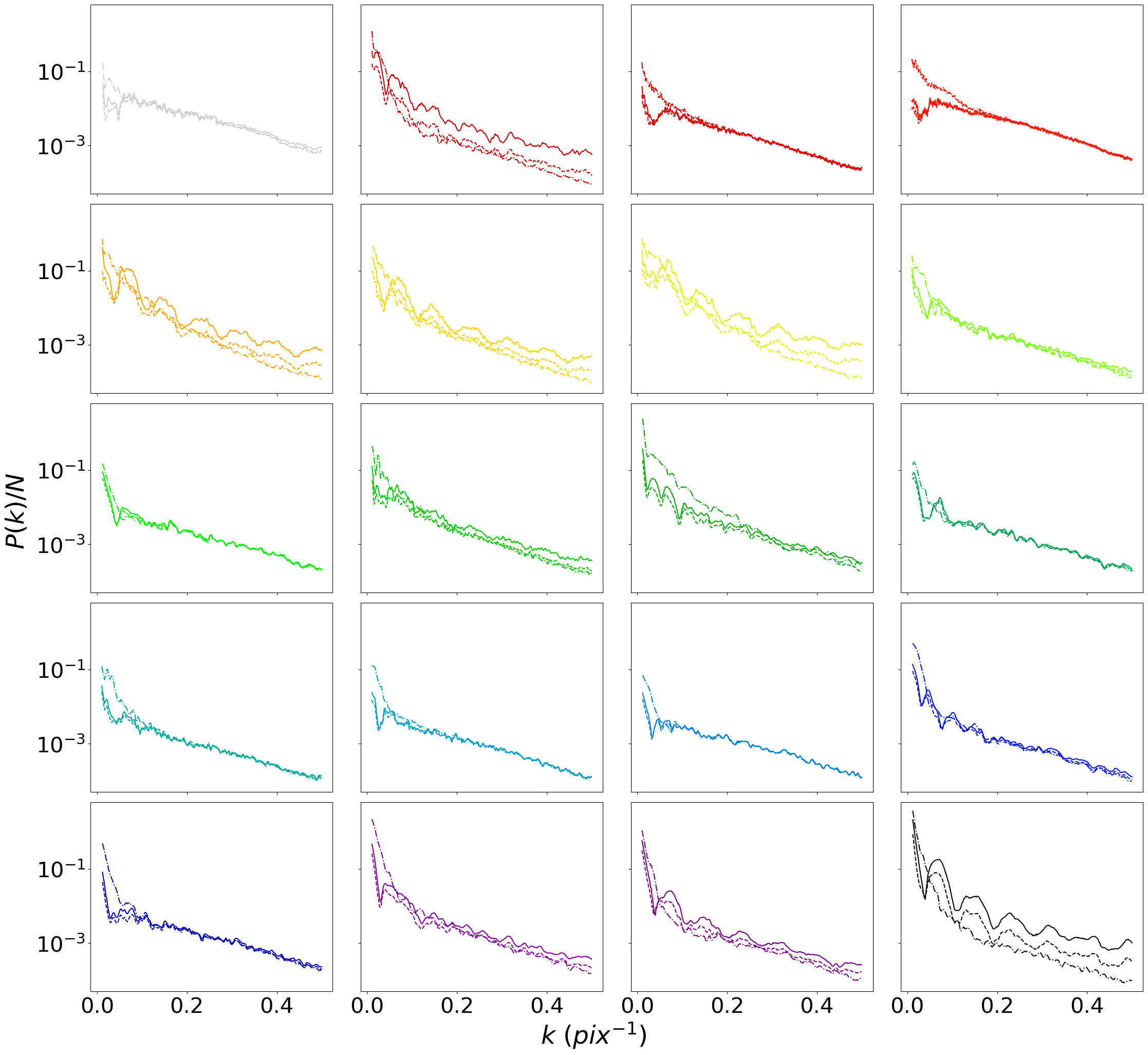}
    \caption{NRI power spectra derived using the normal automatic mask, smoothed edge automatic mask, and smoothed automatic mask. Note that the systems with the largest oscillations in the automatic mask are also those with the largest extra-{\color{black}Gaussian} components in their pixel distribution (see Fig.~\ref{NRI histogram figure}). \textcolor{black}{Galaxies are ordered as in Fig.~\ref{NRI histogram figure}.}}
    \label{comparing power spectra for alternate masking figure}
\end{figure*}

\begin{figure*}
    \centering
    \includegraphics[width=\linewidth]{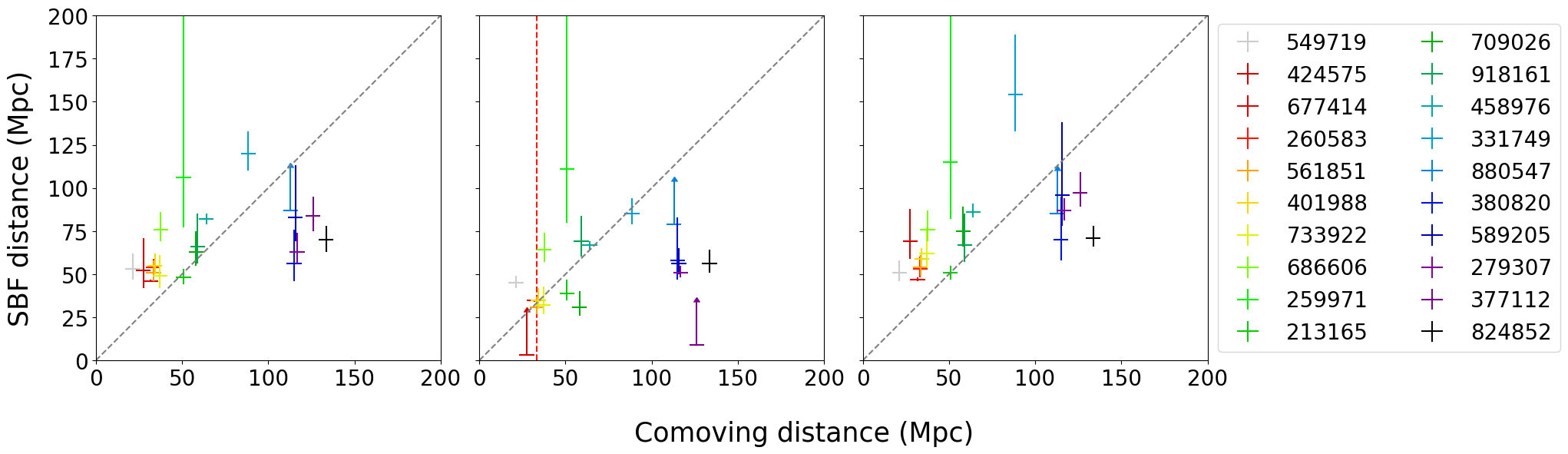}
    \caption{SBF distance estimates derived using the normal automatic mask (left panel), the smoothed-edge automatic mask (centre panel), and the smoothed automatic mask (right panel). A vertical dashed line indicates an object for which SBF were not detected successfully.}
    \label{distance results for alternate masking figure}
\end{figure*}

The first smoothed mask was effective at removing the oscillations in the power spectra, while the smoothed edge mask still showed some oscillations (Fig.~\ref{comparing power spectra for alternate masking figure}). The smoothed mask produced SBF distance results closer to the proper distance, and no longer consistently overestimates the distance of nearby objects (see Fig.~\ref{distance results for alternate masking figure}).
We consider this result is misleading, however, as the extra power available to decrease the SBF distance estimate is clearly related to point sources, and not genuine SBFs. Given the limitations of these versions of the automatic mask, for our fiducial method we chose to use the histogram mask.

%Figure C3

%Figure C4

\section{Variations on the Bicubic Spline Model}

\begin{figure}
    \centering
    \includegraphics[width=0.7\linewidth]{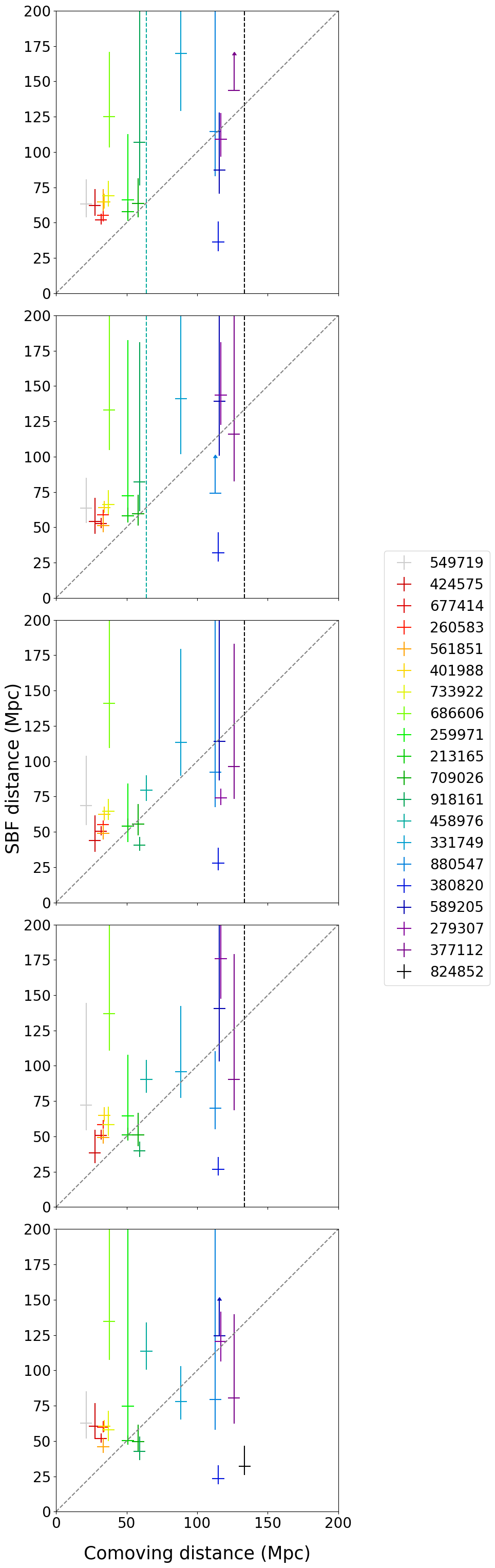}
    \caption{SBF distance estimates derived using the bicubic spline model. From top to bottom, the pixel grid sizes are 20x20, 25x25, 30x30, 35x35, and 40x40 corresponding to a single pixel on the lower resolution image.}
    \label{bicubic spline dsbf vs dc normal mask figure}
\end{figure}

When creating the bicubic spline models, we tested different scales for the low-resolution image, binning together 20x20, 25x25, 30x30, 35x35, and 40x40 pixel areas. Using the same choices as our fiducial method, but the bicubic spline model for the smooth light distribution, we obtain the results given in Fig.~\ref{bicubic spline dsbf vs dc normal mask figure}. The results are generally similar for all the resolutions considered. The nearest galaxies tend to be clustered at similar SBF distances, and the results are less correlated with proper distance compared to Fig.~\ref{dsbf vs dz figure}.

\begin{figure}
    \centering
    \includegraphics[width=0.7\linewidth]{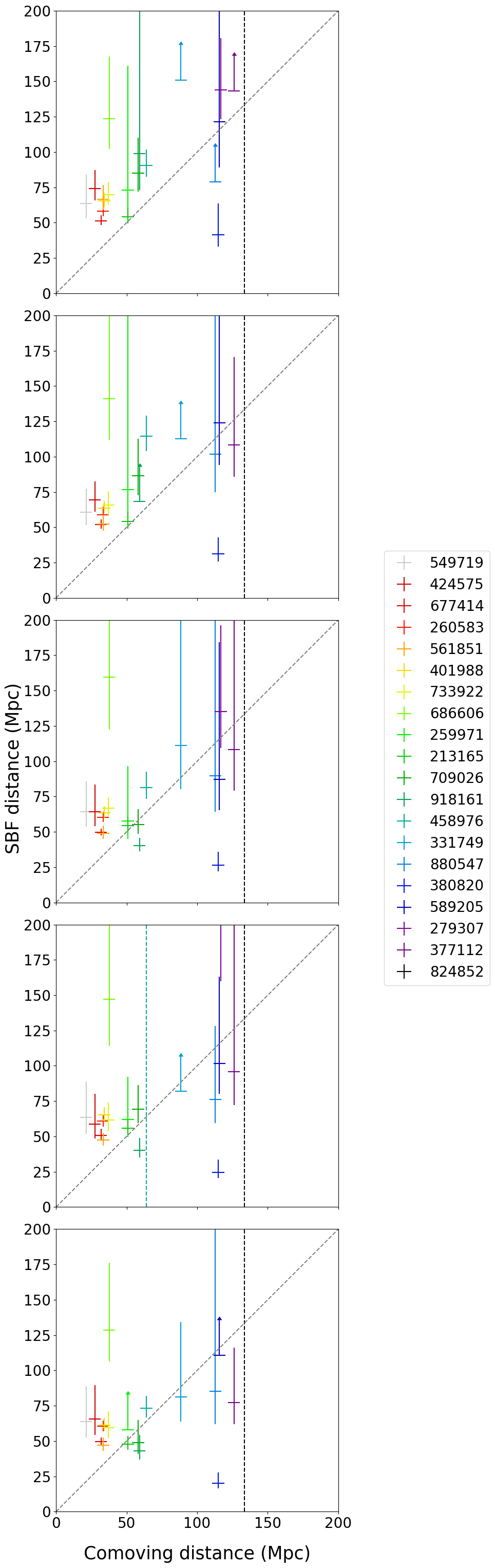}
    \caption{SBF distance estimates derived using the bicubic spline model and the expanded manual mask. From top to bottom, the pixel grid sizes are as in Fig.~\ref{bicubic spline dsbf vs dc normal mask figure}.}
    \label{bicubic spline dsbf vs dc extra mask figure}
\end{figure}

As described in Section \ref{bicubic spline model section}, we also tested using an expanded manual mask to remove pixels contaminated by the mask when producing the lower resolution image. Fig.~\ref{bicubic spline dsbf vs dc extra mask figure} shows the results for various resolutions, with this additional masking. The additional masking reduces the uncertainties in some cases, but generally the results are similar to those in Fig.~\ref{bicubic spline dsbf vs dc normal mask figure}. Since the SBF distances show less correlation with proper distance, relative to our fiducial method, and the NRI histograms are less {\color{black}Gaussian}, as described in Section \ref{model tests section}, we have not adopted the bicubic spline modelling method in our main analysis.

%Figure D1

%Figure D2

%%%%%%%%%%%%%%%%%%%%%%%%%%%%%%%%%%%%%%%%%%%%%%%%%%

\bsp % typesetting comment
\label{lastpage}
\end{document}